\documentclass[floatfix,aps,prd,twocolumn]{revtex4-2}
\usepackage[T1]{fontenc} 
\usepackage{mathptmx}
\usepackage{pslatex}
\usepackage{amsmath}
\usepackage{txfonts}
\usepackage{graphicx}
\usepackage{cancel}
\usepackage{amssymb}
\usepackage{color}

\graphicspath{{./plots/}}

\newcommand{\Refs}[1]{{Refs.~\cite{#1}}}
\renewcommand{\Ref}[1]{{Ref.~\cite{#1}}}
\newcommand{\Eq}[1]{{Eq.~(\ref{#1})}}
\newcommand{\Fig}[1]{{Fig.~\ref{#1}}}
\newcommand{\Tab}[1]{{Tab.~\ref{#1}}}
\def\tbar{{\bar t}}

\begin{document}
\title{Determination of the top-quark mass from top-quark pair events
  with the matrix element method at next-to-leading order: Potential
  and prospects.}

\author{Till Martini}
\thanks{Work on this article was conducted while employed at
  Humboldt-Universität zu Berlin, Institut für Physik, Berlin,
  Germany}
\email{Till.Martini@physik.hu-berlin.de}
\affiliation{Fraunhofer Zentrum SIRIOS, Fraunhofer Institute for
  High-Speed Dynamics EMI, Berlin, Germany}

\author{Turan Nuraliyev}
\email{Turan.Nuraliyev@physik.hu-berlin.de}
 \author{Peter Uwer}
\email{Peter.Uwer@physik.hu-berlin.de}
\affiliation{Humboldt-Universit{\"a}t zu Berlin, Institut f{\"u}r Physik,
  Newtonstra{\ss}e 15, 12489 Berlin, Germany}
\begin{abstract}
  In 2004 the matrix element method was used in a pioneering work by
  the Tevatron experiment D$\cancel{0}$ to determine the top-quark
  mass from a handful of events.  Since then the method has been
  matured into a powerful analysis tool. While the first applications
  were restricted to leading-order accuracy, in the meantime also the
  extension to next-to-leading order (NLO) accuracy has been
  studied. In this article we explore the potential of the matrix
  element method at NLO to determine the top-quark mass using events
  with pair-produced top quarks. We simulate a toy experiment by
  generating unweighted events with a fixed input mass and apply the
  matrix element method to construct an estimator for the top-quark
  mass. Two different setups are investigated: unweighted events
  obtained from the fixed-order cross section at NLO accuracy as well
  as events obtained using \texttt{POWHEG} matched to a parton
  shower. The latter lead to a more realistic simulation and allow to
  study the impact of higher-order corrections as well as the
  robustness of the approach. We find that the matrix element method
  in NLO accuracy leads to a significant reduction of the theoretical
  uncertainties compared to leading order. In view of the high
  luminosity phase of the LHC, this observation is especially relevant
  in analyses which are no longer dominated by statistical
  uncertainties.
 \end{abstract}

\maketitle
\flushbottom

\section{Introduction}
Regarding experimental as well as theoretical progress, hadronic
top-quark pair production has evolved into one of the flagship
processes at the LHC. This development is propelled by the expectation
of the top quark to play a prominent role in extensions of the
Standard Model due to it being by far the heaviest of the elementary
particles with a life time significantly shorter than the time scale
of hadronization. The high production rate of top-quark pairs at the
LHC as well as onward advances in experimental data taking enable for
ever-decreasing statistical and systematic uncertainties in the
recorded data. In order to make optimal use of this fact in
experimental analyses, the employed theoretical predictions are
required to keep up in terms of uncertainties.

The next-to-leading order QCD corrections for top-quark pair
production have been calculated for the spin independent case more
then 30 years ago
\cite{Nason:1987xz,Nason:1989zy,Beenakker:1988bq,Beenakker:1990maa}.
Later, also the spin dependent cross sections were evaluated at NLO
accuracy in QCD \cite{Bernreuther:2004jv,Melnikov:2009dn}.  In a
series of ground breaking articles also the next-to-next-to-leading
order QCD corrections were calculated
\cite{Czakon:2013goa,Czakon:2015owf,Czakon:2016ckf,Catani:2019iny,Catani:2019hip}. Furthermore,
beyond fixed order also the resummation of soft-gluon corrections has
been studied in great detail
(\cite{Beneke:2009ye,Czakon:2009zw,Beneke:2011mq,Cacciari:2011hy,%
  Kidonakis:2012rm,Ferroglia:2012ku,Ferroglia:2013awa,Czakon:2018nun}).
In addition to QCD corrections also weak and QED corrections have been
calculated
\cite{Beenakker:1993yr,Bernreuther:2006vg,Kuhn:2006vh,Moretti:2006nf,
  Pagani:2016caq}. In
summary, many detailed theoretical predictions for top-quark pair
production are available.  However, these might not be readily
applicable in the experimental analysis. It is thus important to put
more effort in improving the interface between experiment and theory
to make optimal use of the increasing precision reached in both
fields.

Multivariate analysis methods like the matrix element method (MEM),
turn out to be particularly useful in making optimal use of the
theoretical predictions. The MEM requires the calculation of event
weights in terms of differential cross sections and is thus often
formulated at lower-order accuracy only. At leading order (LO), the
MEM has been established as a powerful analysis tool for both signal
searches as well as parameter inference by virtue of its optimal
utilization of the information content of the available
data. Typically, the impact of higher-order QCD corrections on
theoretical predictions can be significant while often simultaneously
decreasing the theoretical uncertainties. In the quest for accuracy
and precision to match experimental achievements, the MEM at
next-to-leading order (NLO) represents a promising remedy. But when
taking higher-order corrections into account, the calculation of event
weights constitutes a non-trivial task due to the intricate
combination of virtual and real contributions to obtain meaningful
finite results. The problem of extending the MEM beyond the Born
approximation has been solved in the past by introducing modified jet
algorithms on the one hand or sensible event definitions on the other
hand (\cite{Martini:2015fsa,Martini:2017ydu,Kraus:2019qoq}). At the
same time, the application of the MEM at NLO has been demonstrated for
top-quark mass extraction from simulated single top-quark events
(\cite{Martini:2015fsa,Martini:2017ydu,Kraus:2019qoq}) as well as
anomalous coupling parameter determination from simulated Higgs boson
events in association with a single top quark
(\cite{Kraus:2019myc}). Additionally, the effects of a parton shower
applied to simulated single top-quark data has been investigated with
the MEM at NLO (\cite{Kraus:2019qoq}). In this work, we present the
application of the MEM at NLO to top-quark pair production at the
LHC. In contrast to the electroweak production mechanism of single top
quarks studied before, top-quark pair production is QCD-induced at LO
already with the two production channels of quark-antiquark
annihilation and gluon-gluon fusion constituting the dominant source
of top quarks at the LHC. Given the aforementioned prominent role of
top-quark pair production in both experimental as well as theoretical
advances at the LHC, it represents an ideal example to study
higher-order effects within the MEM. Furthermore, in view of the
ongoing progress in top-quark mass measurements, the MEM at NLO
accuracy could be an interesting alternative to existing approaches.

The paper is structured as
follows. In section~\ref{sec:ttbarLHC} the NLO QCD calculation of the
differential cross section for top-quark pair production with the
phase space slicing method and the subsequent generation of unweighted
events are briefly reviewed. Section~\ref{sec:appl} focuses on the
application of the MEM to the generated events. To study parton shower
effects, events generated with \texttt{POWHEG+Pythia}
\cite{Nason:2004rx,Sjostrand:2006za,Frixione:2007vw,Frixione:2007nw,
  Alioli:2010xd}
are also analysed. The conclusions are presented in section~\ref{sec:concl}.

\section{Top-quark pair production at the LHC\label{sec:ttbarLHC}}
\subsection{Implementing the NLO prediction with the phase space slicing method}
The MEM at NLO as presented in
\cite{Martini:2015fsa,Martini:2017ydu,Kraus:2019qoq} requires the
cross-section calculation at NLO to be carried out using the phase
space slicing method \cite{Giele:1993dj}. The respective calculation
is available in the literature \cite{Bernreuther:2004jv}. Thus, in
this section we only give a brief review of the important aspects of
the calculation and present the validation for the choice of the
\textit{slicing parameter}. In the phase space slicing method, the
cross-section prediction at NLO accuracy ${\rm d}\sigma_{\rm NLO}$ is
formed of two contributions: First, the so-called \textit{hard part}
${\rm d}\sigma_{\rm Hard}$ is just the matrix element for the real
corrections evaluated for phase space points where all partons are
resolved, that is the additional parton is neither collinear to the
incoming partons nor soft. Second, a Born-like part is comprised of
the Born contribution ${\rm d}\sigma_{\rm LO}$, the virtual
corrections ${\rm d}\sigma_{\rm virtual}$ (taken from
\Ref{Badger:2011yu}) as well as the so-called \textit{soft} and
\textit{collinear parts} ${\rm d}\sigma_{\rm soft/coll.}$ stemming
from approximated real corrections integrated over phase space regions
in which the additional parton is unresolved. The separation of the
phase space for the real corrections into resolved and unresolved
regions is mediated by the so-called slicing parameter $x_{\rm min}$
which acts as a scale to separate the two. In the unresolved regions,
well-known factorization properties of QCD real corrections can be
employed allowing to analytically integrate over the additional
radiation in the singular limits in an approximate way thereby
reducing the respective phase space to Born-like kinematics. The
divergences of these integrations can be regularized within
dimensional regularization leading to poles in the dimensional shift
away from four space-time dimensions. The outcome can be
combined with the virtual contributions to cancel the respective poles
from the loop integration and yield finite results according to the
Kinoshita-Lee-Nauenberg theorem
(\cite{Kinoshita:1962ur,Lee:1964is}). Since the real corrections are
approximated in the unresolved (singular) regions, the result is only
accurate up to deviations proportional to the slicing parameter
$x_{\rm min}$:
\begin{equation}\label{eq:NLOpss}
{\rm d}\sigma_{\rm NLO}\;=\;{\rm d}\sigma_{\rm Hard}\;+ \;{\rm d}\sigma_{\rm LO}\; +\; {\rm d}\sigma_{\rm virtual}\; +\; {\rm d}\sigma_{\rm soft/coll.}\;+\; {\cal O}(x_{\rm min})\;.
\end{equation}
Additionally, the separation of the real phase space in terms of the
slicing parameter introduces logarithmic dependencies of the hard and
soft/collinear contributions on $x_{\rm min}$ which cancel in the
sum. However, when numerically integrating over the finite hard
contribution, these logarithms can lead to numerical instabilities if
$x_{\rm min}$ is chosen too small. Hence, the value of $x_{\rm min}$
has to be chosen as a compromise between numerical stability and the
demand that the deviation in \Eq{eq:NLOpss} is negligible
compared to the statistical uncertainties of the total cross section
as well as distributions calculated at NLO accuracy.
\begin{figure}
\includegraphics[width=\columnwidth]{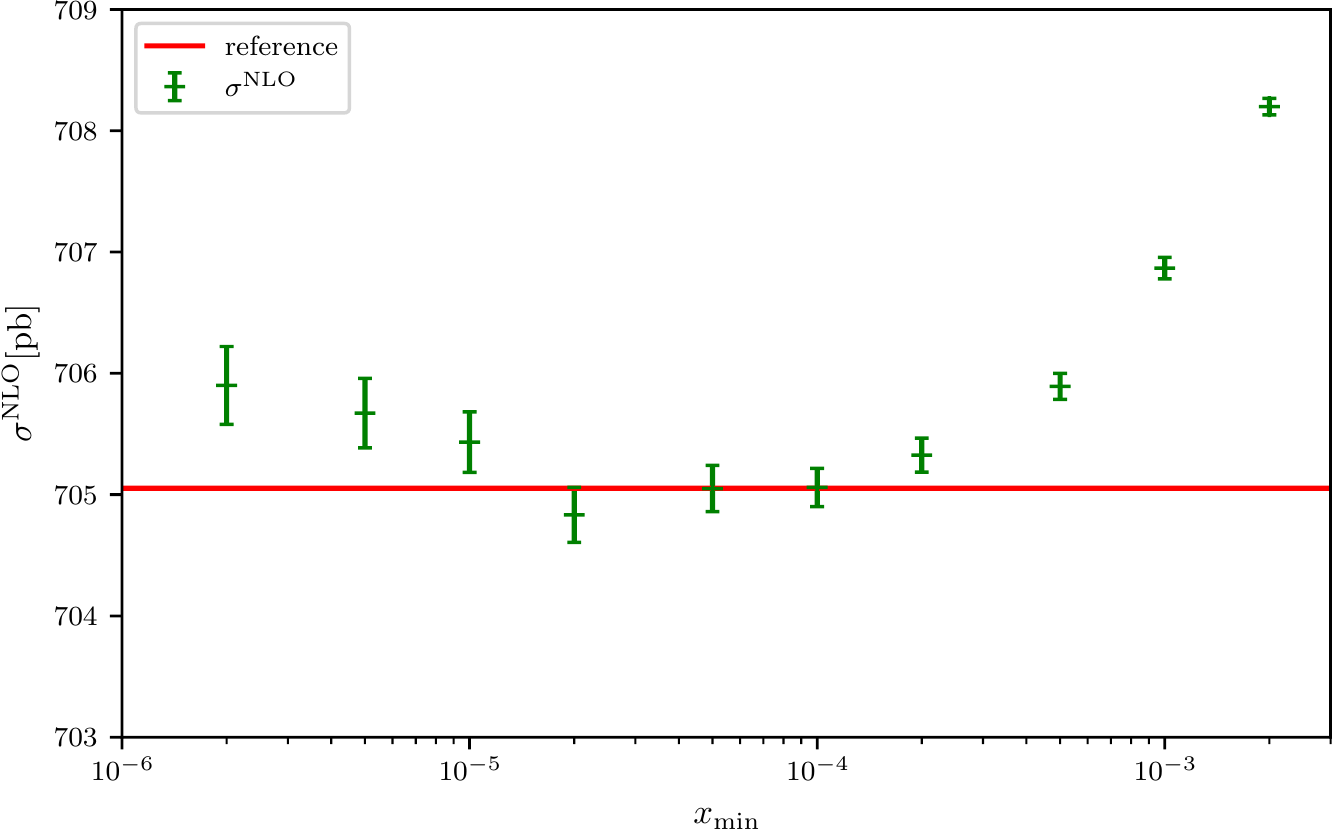}
\caption{Phase space slicing parameter (in-)dependence of the total
  cross section predicted at NLO accuracy. The red line shows the
  reference value taken from \texttt{HATHOR}
  \cite{Aliev:2010zk}. \label{fig:pssvalid-total}}
\end{figure}
\begin{figure}
\includegraphics[width=\columnwidth]{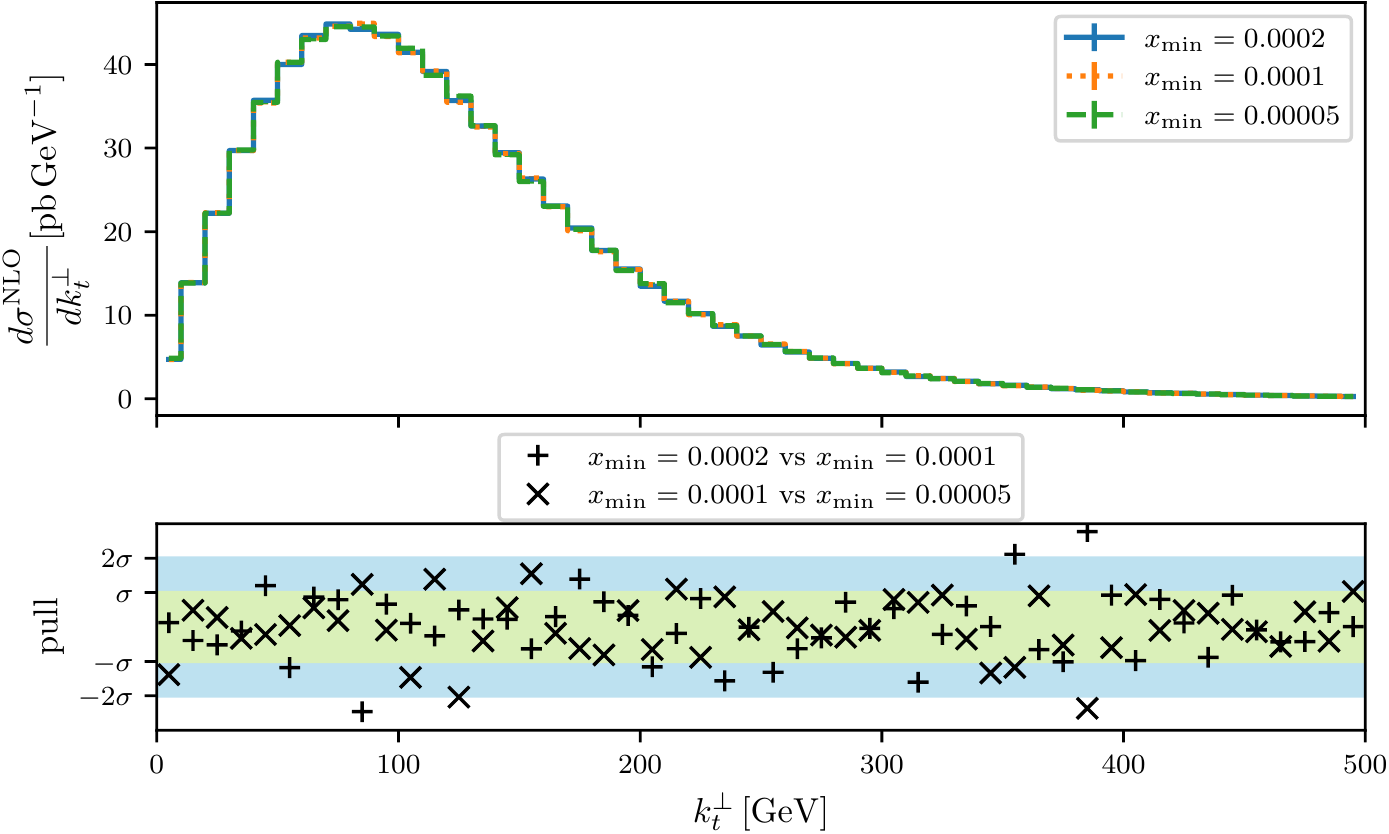}
\caption{Phase space slicing parameter (in-)dependence of the
  top-quark transverse momentum $k_t^\perp$ predicted at NLO
  accuracy.\label{fig:pssvalid-diff}}
\end{figure}
\Fig{fig:pssvalid-total} shows NLO predictions for the total cross
section of top-quark pair production for different values of the
slicing parameter $x_{\rm min}$. The total cross section as the sum of
Born, virtual and real contributions in \Fig{fig:pssvalid-total} is
indeed finite. However, it shows a systematic deviation from the
reference value taken from \texttt{HATHOR} \cite{Aliev:2010zk} for
values $x_{\rm min}\gtrapprox 2\times10^{-3}$ while for values
$x_{\rm min}\lessapprox 5\times10^{-6}$ numerical instabilities
dominate. Accordingly, a value of $x_{\rm min}= 10^{-4}$ is chosen.
As an example of a differential distribution the top-quark transverse
momentum calculated at NLO accuracy is shown for three different
choices of $x_{\rm min}$ in \Fig{fig:pssvalid-diff}. In the lower plot
we show for different choices of $x_{\rm min}$ the differences in
units of the statistical uncertainties. We conclude that all three
choices lead to coherent predictions justifying the choice
$x_{\rm min}= 10^{-4}$. In addition to the top-quark transverse
momentum this has been checked also for the top-quark energy
distribution and the rapidity distribution. Furthermore, the
distributions calculated here have been cross checked with results
from \texttt{madgraph5 aMC@NLO} \cite{Alwall:2014hca}. The comparison
is shown in appendix \ref{AdditionaCrossChecks}, \Fig{fig:diffvalid1}
and \Fig{fig:diffvalid2}.
\begin{figure}
\begin{center}
\includegraphics[width=\columnwidth]{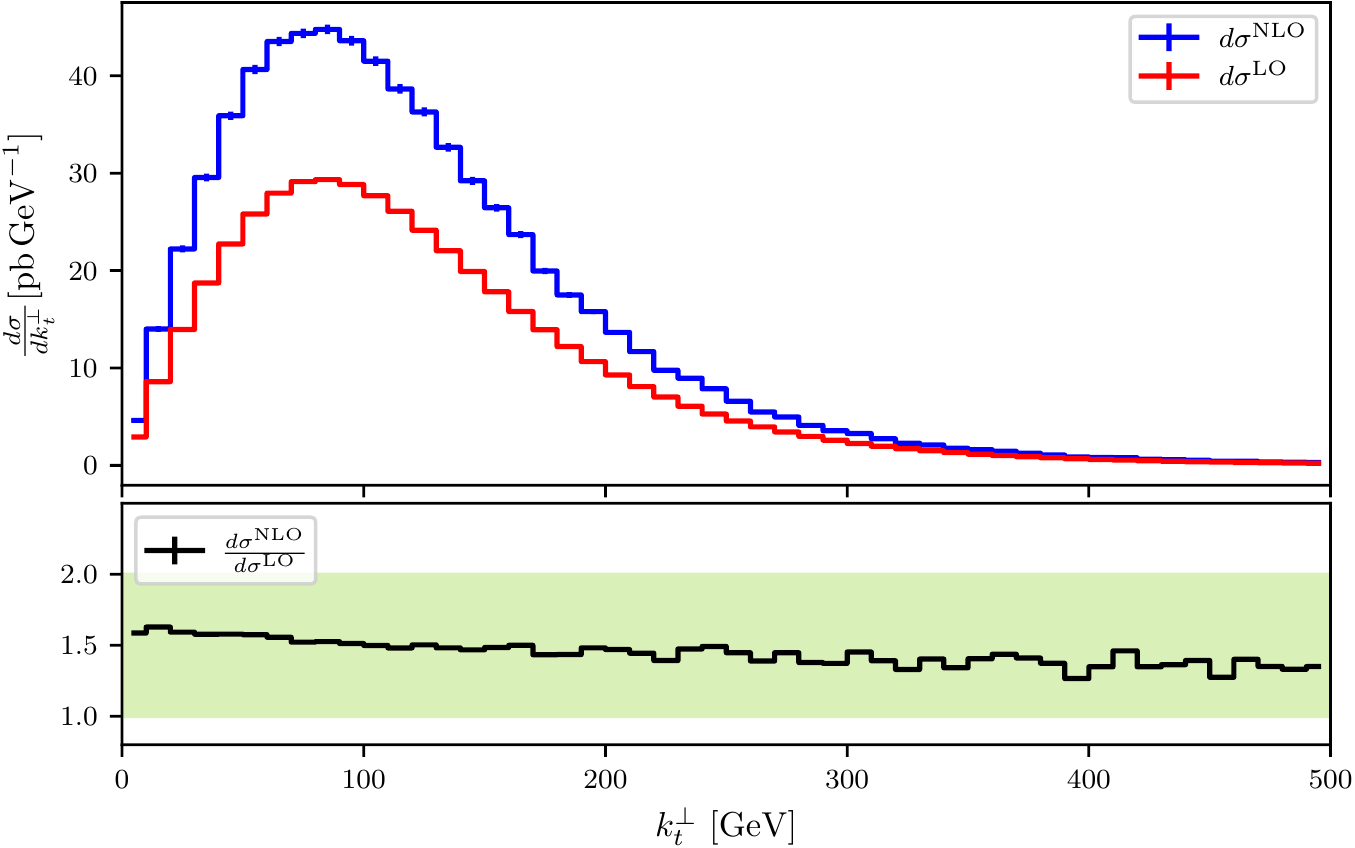}\\
\includegraphics[width=\columnwidth]{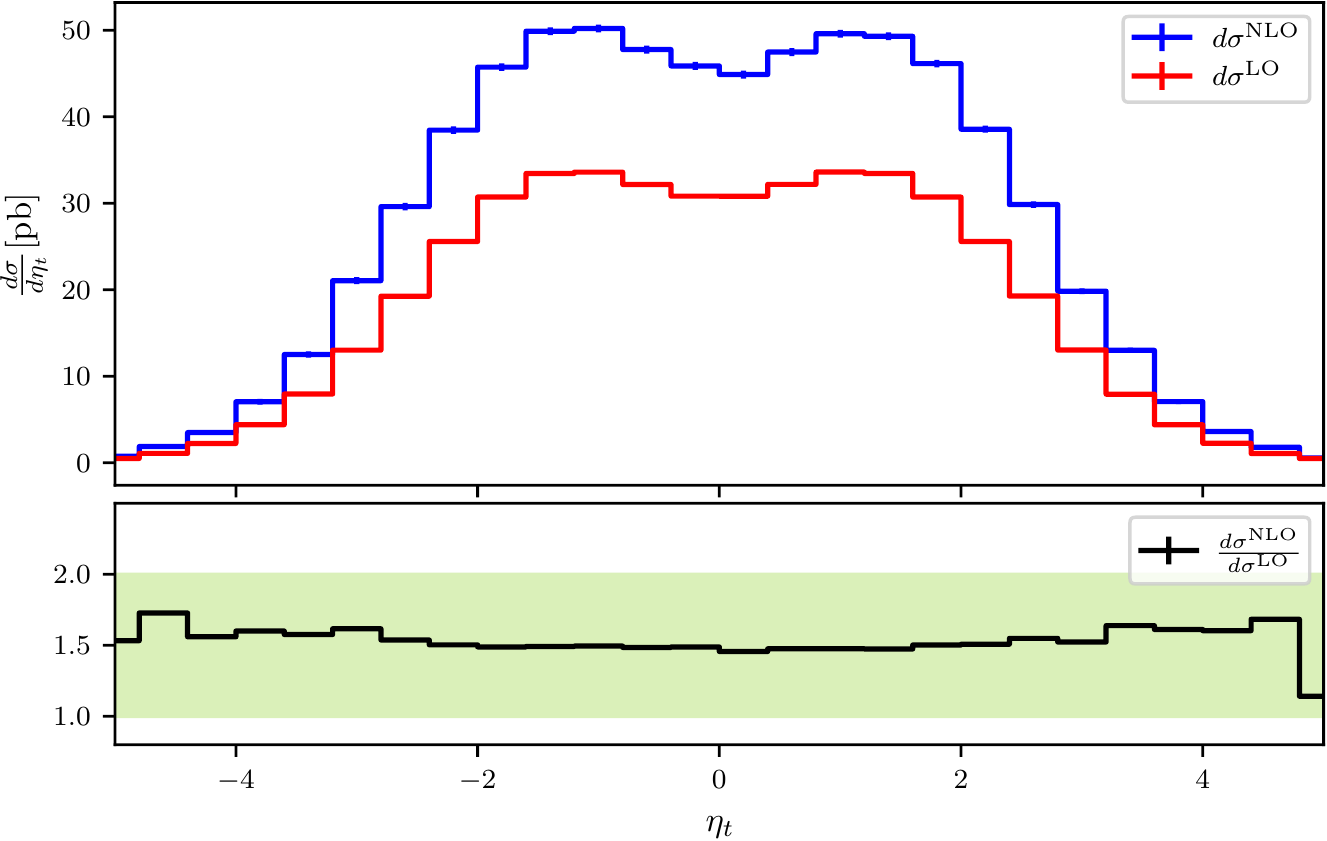}
\end{center}
\caption{Differential distributions together with the
  respective $k$-factors.\label{fig:kfacs1}}
\end{figure}
The impact of the NLO corrections on kinematic distributions is
displayed in \Fig{fig:kfacs1} where NLO and LO predictions for
kinematic distributions are compared and their ratios (the
\textit{k-factor}) are shown at the bottom of the plots. Results for
further distributions are shown in \Fig{fig:kfacs2} in
appendix~\ref{AdditionaCrossChecks}. As can be seen from the rather
constant k-factors, the NLO corrections only mildly affect the shapes
of the kinematic distributions. However, the NLO corrections lead to a
significant increase of the cross sections by a factor of roughly
$1.5$.
\begin{figure}
\begin{center}
\includegraphics[width=\columnwidth]{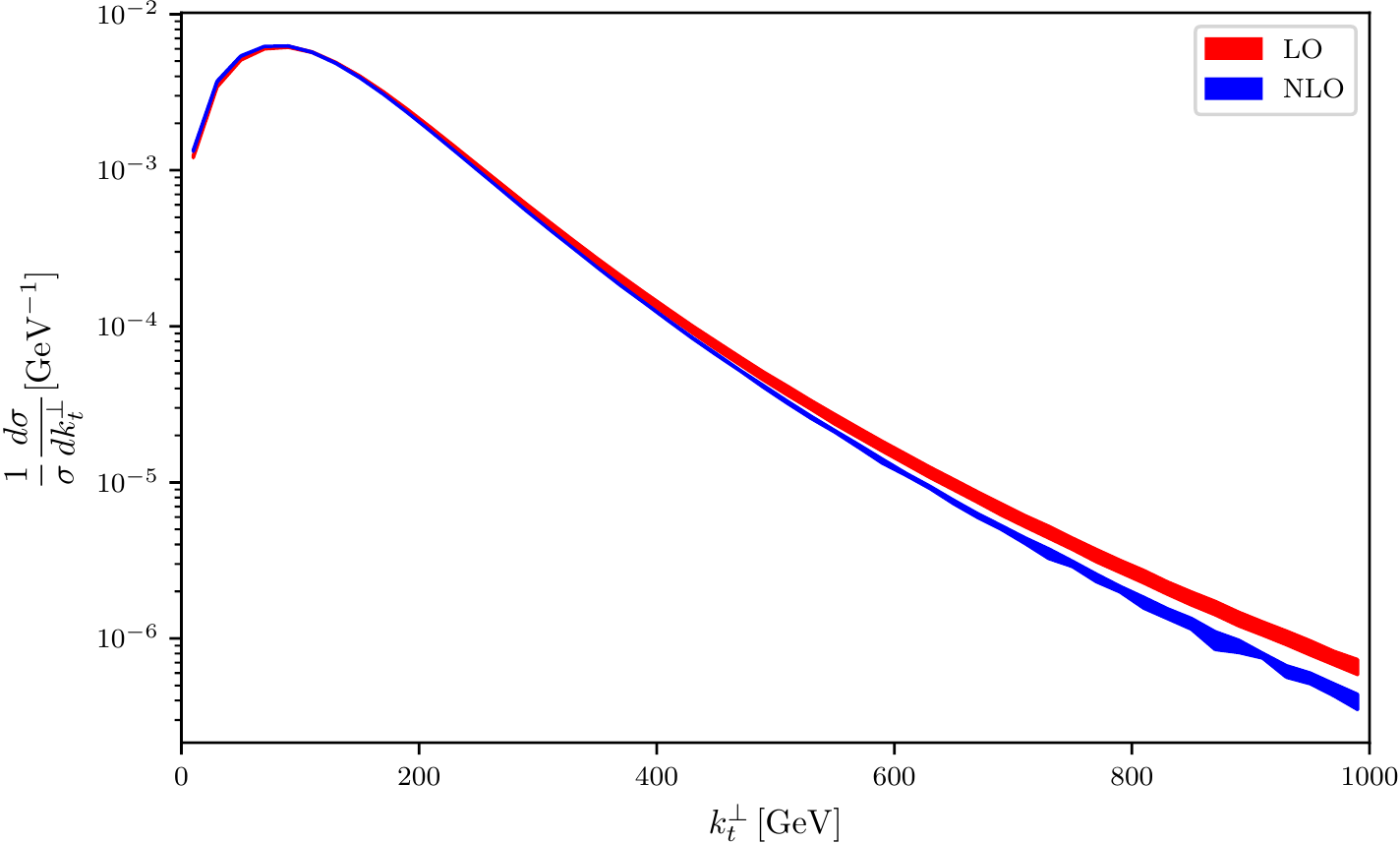}\\
\includegraphics[width=\columnwidth]{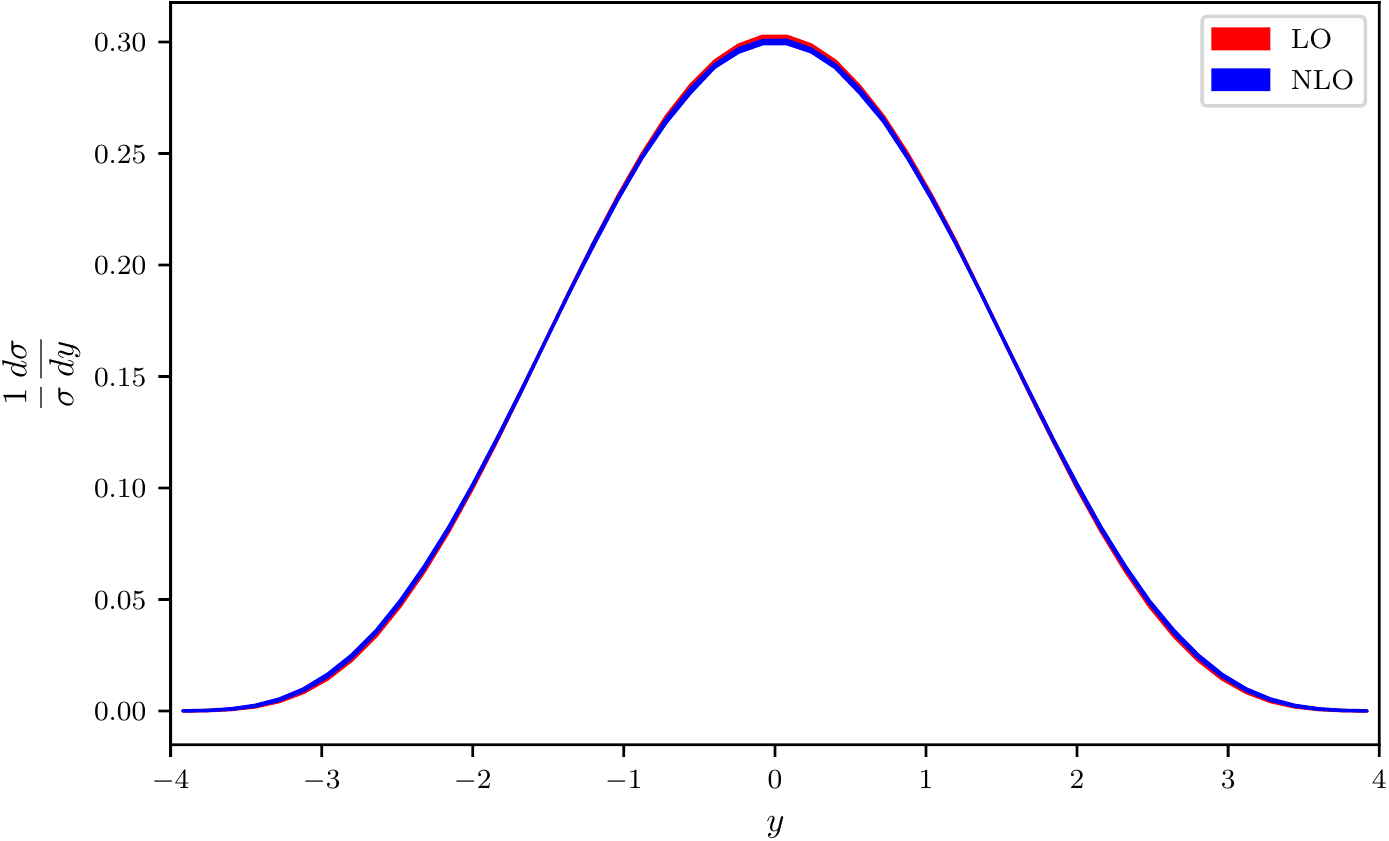}
\end{center}
\caption{Effect of scale variations on the shapes of kinematic
  distributions of the top quark.\label{fig:scalevar}
  }
\end{figure}
In \Fig{fig:scalevar} the impact of variations of the factorization
scale $\mu_{\rm F}$ and renormalization scale $\mu_{\rm R}$ by a
factor of $2$ as a means to estimate the effect of un-calculated
higher orders are illustrated for the shapes of two representative
kinematic distributions of the top quark. For moderate energy scales,
one observes a significant reduction of the impact of the scale
variation.

\subsection{Unweighted event generation}
From the calculation of the cross section at NLO accuracy outlined in
the previous section, event weights can be calculated which can be used to
generate unweighted events which are distributed according to the NLO
cross section. As described in \Ref{Kraus:2019qoq}, a sensible
event definition is mandatory for obtaining meaningful event weights
at NLO accuracy. In particular, the event definition must not fix the
invariant masses or the overall transverse momentum of the final-state
objects. For top-quark pair production, we define events $\vec{x}$
in terms of the transverse momentum $k_t^\perp$, azimuthal angle
$\phi_t$ and pseudo rapidity $\eta_t$ of the top quark as well as the
pseudo rapidity of the antitop quark $\eta_\tbar$:
\begin{equation}
\vec{x}=(k_t^\perp,\phi_t,\eta_t,\eta_\tbar)\;.
\end{equation}
The two-particle Born phase space as well as the three-particle phase
space for the real radiation can be parameterized in terms of these
variables
\begin{eqnarray}
{\rm
  d}R_2&=&\frac{{k_t^\perp}^3\;\cosh{\eta_t}\;\cosh{\eta_\tbar}}{8\pi^2\;E_t\;E_\tbar\;
           s_{\rm had}}\;{\rm d}k_t^\perp\;{\rm d}\phi_t\;{\rm d}\eta_t\;{\rm d}\eta_\tbar\;,\\
{\rm
  d}R_3&=&\frac{{k_t^\perp}^2\;{k_\tbar^\perp}\;{k_3^\perp}^2\;\cosh{\eta_t}\;\cosh{\eta_\tbar}\;
           \cosh{\eta_3}}{128\pi^5\;E_t\;E_\tbar\;E_3\;s_{\rm had}}\;\nonumber \\ 
   &\times&        
           {\rm
           d}k_t^\perp\;
           {\rm d}\phi_t\;{\rm d}\eta_t\;{\rm d}\eta_\tbar\;{\rm d}k_3^\perp\;{\rm d}\phi_3\;{\rm d}\eta_3\;,
\end{eqnarray}
where $E_i$ ($i=t,\tbar,3$) denotes the energy of particle $i$ and  $s_{\rm had}$ is the
hadronic center-of-mass energy squared. The additional radiation
occurring in the real corrections is parametrized by the transverse
momentum $k_3^\perp$, the azimuthal angle $\phi_3$ and the pseudo
rapidity $\eta_3$ of the radiated parton. These parametrizations allow
together with \Eq{eq:NLOpss} 
to calculate the event weight at NLO accuracy for each event $\vec{x}$  using
\begin{eqnarray}\label{eq:evwgtNLO}
  \nonumber
  \frac{{\rm d}^4\sigma_{\rm NLO}}{{\rm d}k_t^\perp\;{\rm
  d}\phi_t\;{\rm d}\eta_t\;{\rm d}\eta_\tbar} &=&
\;\frac{{\rm d}^4\sigma_{\rm LO}}{{\rm d}k_t^\perp\;{\rm
   d}\phi_t\;{\rm d}\eta_t\;{\rm d}\eta_\tbar}\;
                                              \nonumber \\
  &&\hspace{-2cm}+\int \frac{{\rm d}^7\sigma_{\rm Hard}}{{\rm d}k_t^\perp\;{\rm
      d}\phi_t\;{\rm d}\eta_t\;{\rm d}\eta_\tbar\;{\rm d}k_3^\perp
      \;{\rm d}\phi_3\;{\rm d}\eta_3}\;
      {\rm d}k_3^\perp\;{\rm d}\phi_3\;{\rm d}\eta_3\nonumber \\
  &&\hspace{-2cm}
   +\; \frac{{\rm d}^4\sigma_{\rm virtual}}{{\rm d}k_t^\perp\;{\rm
   d}\phi_t\;{\rm d}\eta_t\;{\rm d}\eta_\tbar}\;
   +\; \frac{{\rm d}^4\sigma_{\rm soft/collinear}}{{\rm d}k_t^\perp\;{\rm d}\phi_t\;{\rm d}\eta_t\;{\rm d}\eta_\tbar}\;.
\end{eqnarray}
\begin{figure}
  \begin{center}
  \includegraphics[width=\columnwidth]{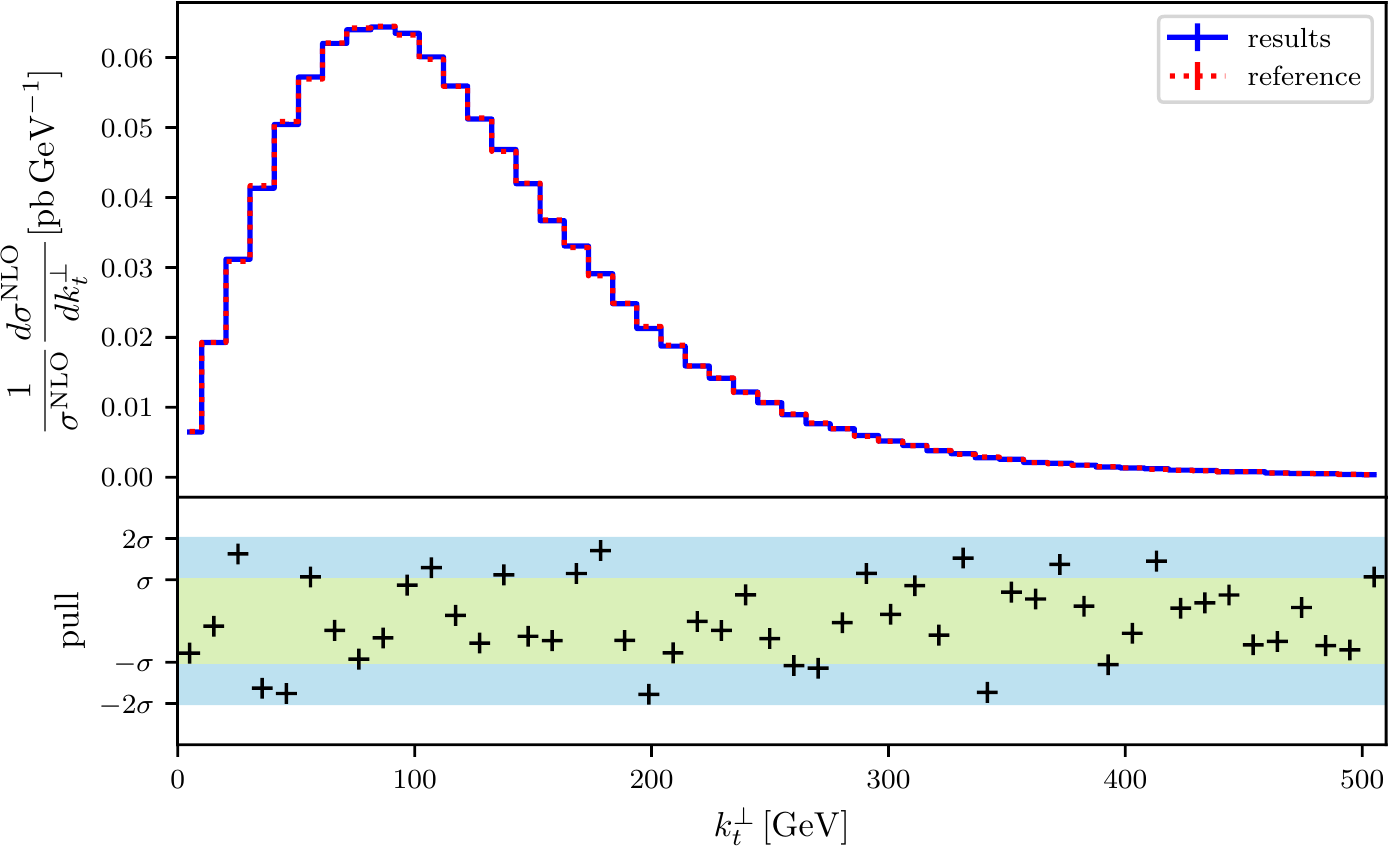}\\
  \includegraphics[width=\columnwidth]{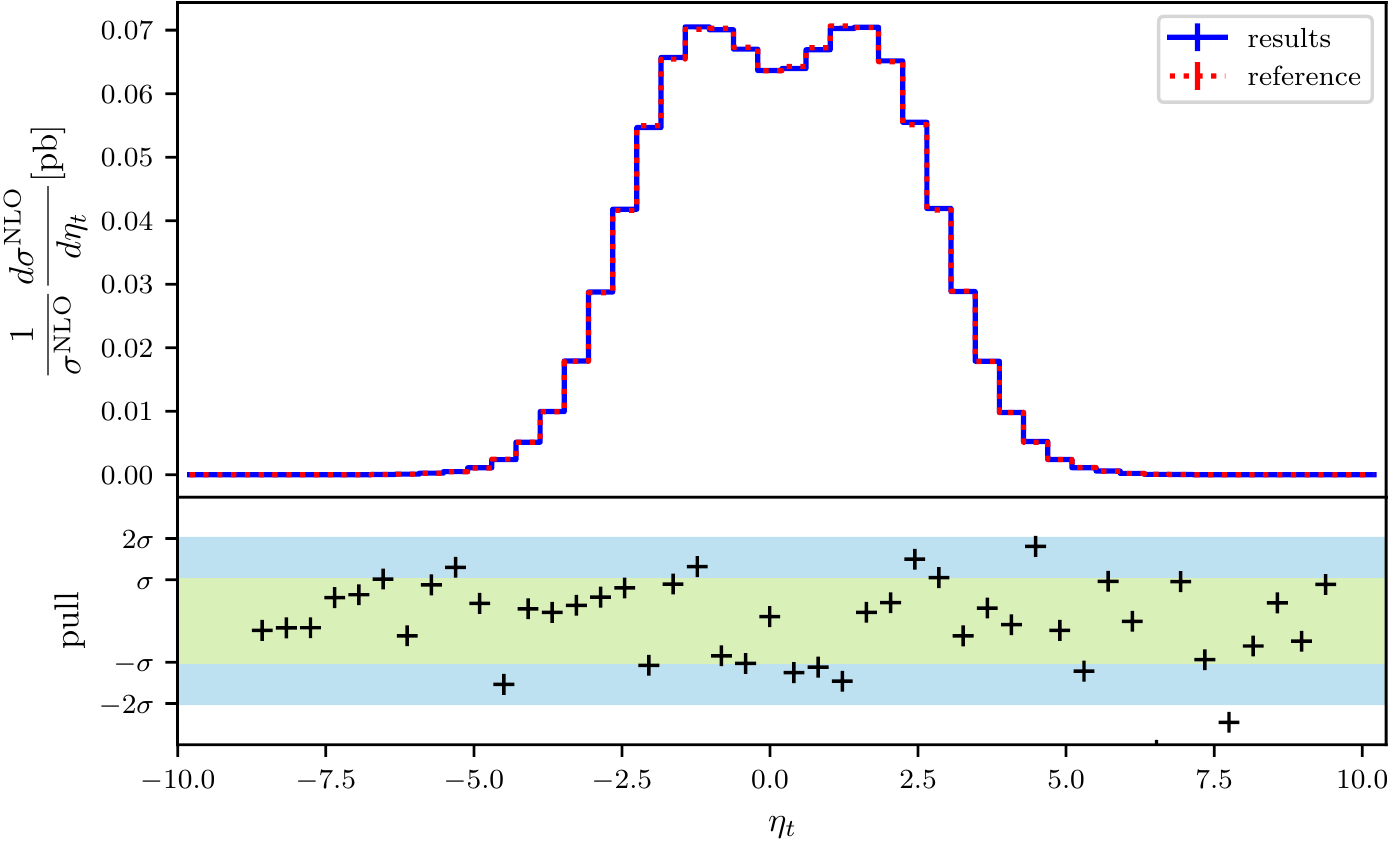}    
  \end{center}
\caption{Validation of the event generation: Comparison of
  differential distributions of the top quark obtained from unweighted
  events with results from \texttt{madgraph5
    aMC@NLO}.\label{fig:evvalid1}}
\end{figure}%
The weights calculated in this way can also be used to generate unweighted
events with, e.g., the \textit{von-Neumann acception-rejection method}
(\cite{vonNeumann1951}). \Fig{fig:evvalid1} shows the distribution of
the unweighted events compared to kinematic distributions obtained
with the \texttt{madgraph5 aMC@NLO} code \cite{Alwall:2014hca}. The
events obtained from the event weights defined in \Eq{eq:evwgtNLO} are
within the uncertainties in perfect agreement with the predictions
obtained using \texttt{madgraph5 aMC@NLO}. In appendix~\ref{AdditionaCrossChecks}, \Fig{fig:evvalid2} we show
in addition the calculation of the $M_{t\bar t}$ and the
$\phi_t$-distribution with the same perfect agreement. The comparison
of the generated unweighted events with the results from \texttt{madgraph5
  aMC@NLO} also serves as a further validation for the choice of the
slicing parameter.

\section{Application: Determination of the top-quark mass using the
  MEM at NLO\label{sec:appl}}
The event weights defined in \Eq{eq:evwgtNLO} can be used in the MEM to
calculate the likelihood at NLO accuracy for a given sample of $N$
events $\{\vec{x}_i\},\;i=1,\ldots,N$:
\begin{equation}\label{eq:likelihood}
{\cal L}\left(\{\vec{x}_i\}\;|\;m_t\right)=\frac{1}{\left(\sigma_{\rm
      NLO}(m_t)\right)^N}\prod\limits_{i=1}^{N}\left.\frac{{\rm
      d}^4\sigma_{\rm NLO}(m_t)}{{\rm d}k_t^\perp\;{\rm d}\phi_t\;{\rm
      d}\eta_t\;{\rm
      d}\eta_\tbar}\right|_{\vec{x}=\vec{x}_i}
\end{equation}
where the dependence of the total and differential cross sections on
the value of the top-quark mass is highlighted{\textemdash}exemplarily
for generic model parameters. Here, the so-called \textit{transfer
  functions}, parametrizing the probability of measuring a certain
signal in the detector given a particular partonic configuration, are
set to delta functions. The transfer functions account for particle
decays, additional radiation as well as detector effects. Thus, this
choice for the transfer functions corresponds to the assumption of a
perfect detector which allows a perfect unfolding from the detector
signals to partonic variables. While for variables related to angles,
setting the transfer function to delta function may give a reasonable
approximation, this is not necessarily true in case of variables
sensitive to energies. In future applications non-trivial transfer
functions should thus be incorporated. This may be done using
invertible neural networks trained to a full simulation as discussed
in great detail in \Ref{Butter:2022vkj}.  This is however beyond the
scope of this work which focuses on exploring the potential of the
method for top-quark mass measurements. Maximizing the likelihood with
respect to the parameter $m_t$ yields an estimator for the top-quark
mass $\hat{m}_t$:
\begin{equation}
\label{Definition-likelihood}
{\cal L}\left(\{\vec{x}_i\}\;|\;\hat{m}_t\right) =
\max_{m_t}\left({\cal L}\left(\{\vec{x}_i\}\;|\;m_t\right)\right)\;.
\end{equation}
Because the event weights in \Eq{eq:likelihood} are normalized to
yield probabilities, the MEM is only sensitive to the shapes of
kinematic distributions but not to the total number of events in the
sample. To also benefit from the information of the total event number
the so-called \textit{extended likelihood} can be used. The extended
likelihood is obtained from the likelihood in
\Eq{Definition-likelihood} by multiplying with the \textit{Poisson}
probability for observing $N$ events when the expected number of
events is given by the total cross section times the integrated
luminosity $L_{\rm int}$ of the collider:
\begin{equation}
  {\cal L}_{\rm ext}\left(\{\vec{x}_i\}\;|\;m_t\right)=
  \frac{\left(\sigma_{\rm NLO}(m_t)\;L_{\rm int}\right)^N}{N!}
  e^{-\sigma_{\rm NLO}(m_t)\;
    L_{\rm int}}\;{\cal L}\left(\{\vec{x}_i\}\;|\;m_t\right).
\end{equation}
\begin{figure}
  \begin{center}
    \includegraphics[width=\columnwidth]{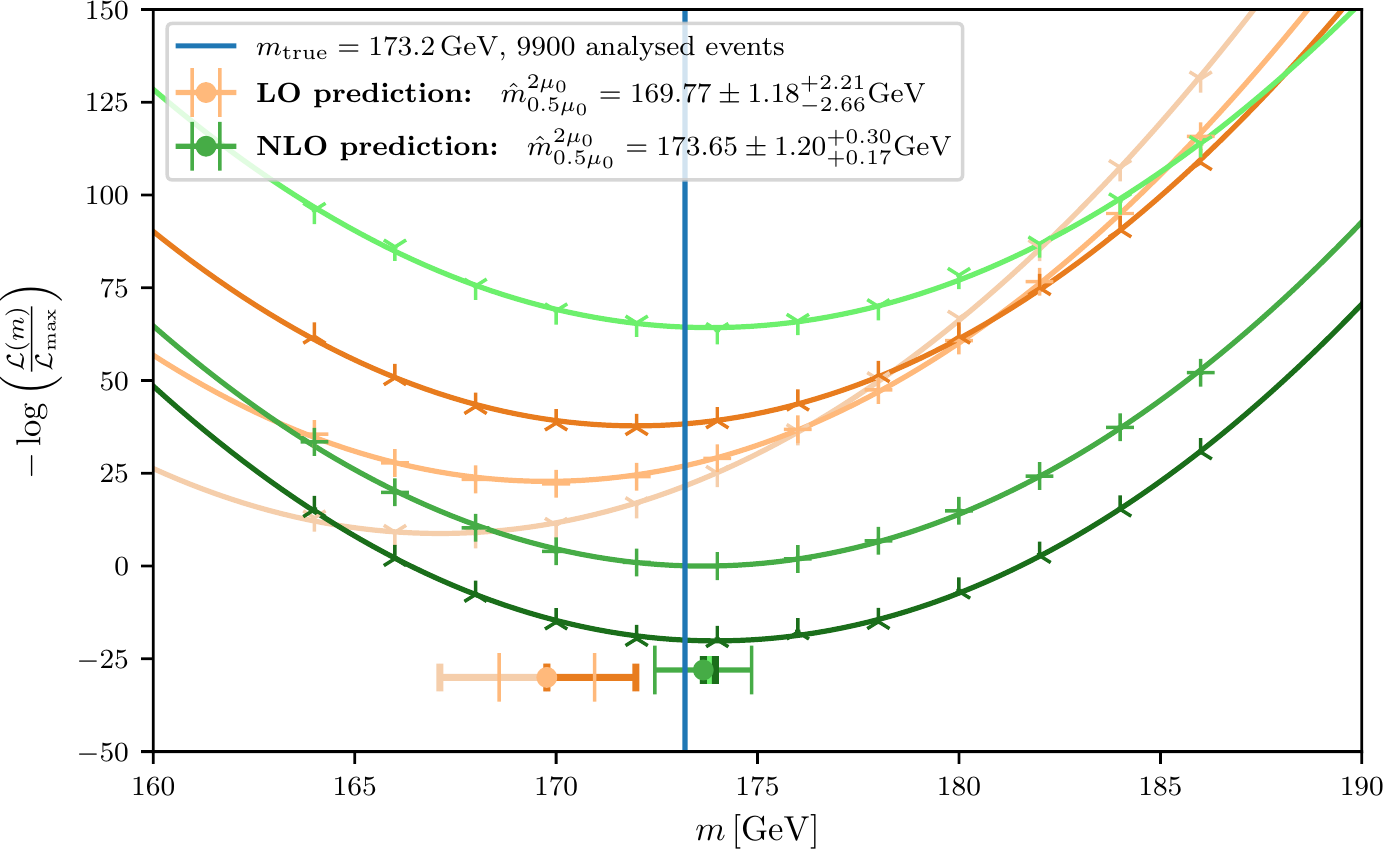}\\
    \includegraphics[width=\columnwidth]{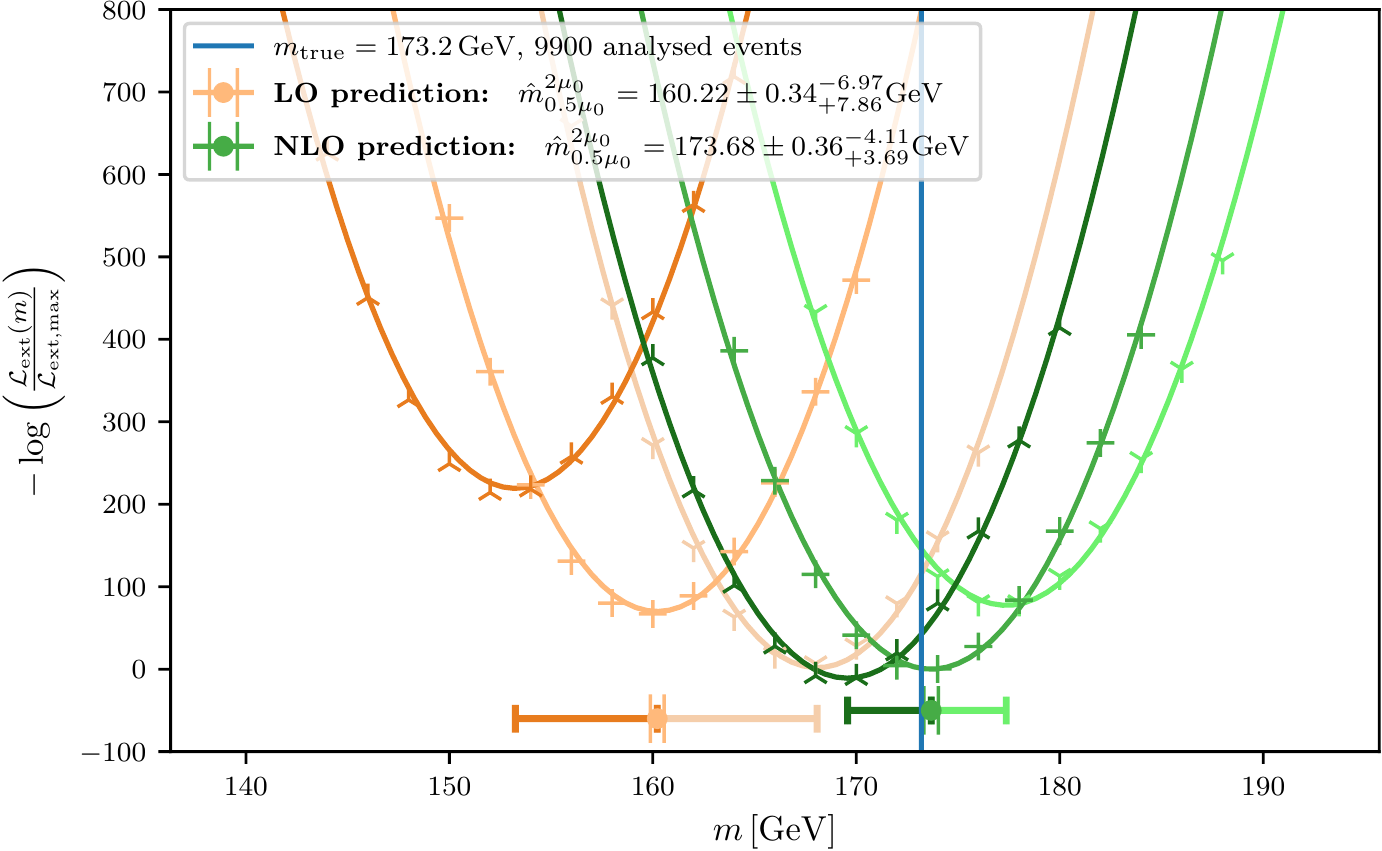}    
  \end{center}
  \caption{Analysis of unweighted events following the fixed-order NLO
    prediction with (extended) likelihoods calculated at LO and NLO
    accuracy. 
    \label{fig:memunwgtev} }
\end{figure}
In \Fig{fig:memunwgtev} we show the likelihood obtained analysing
$9900$ unweighted top-quark pair events distributed according to the
NLO prediction. Likelihood (upper plot) as well as the extended
likelihood (lower plot) have been studied. The green curves correspond
to likelihoods calculated at NLO accuracy using different choices for
the factorization and renormalization scale. The orange curves are
obtained using only LO predictions again for different scale settings
in the likelihood calculation. The analysed events are generated for
an input value of the top-quark mass of
$m_{\mbox{\scriptsize true}}=173.2$~GeV and the scale choice
$\mu_{\rm F}=\mu_{\rm
  R}=\mu_{0}=m_t$.\renewcommand{\arraystretch}{1.3}
\begin{table}
  \centering
\begin{tabular}{|r| r | r |}
\cline{2-3}
 \multicolumn{1}{c|}{}& \multicolumn{2}{|c|}{}\\[-2ex]
 \multicolumn{1}{c|}{}& \multicolumn{2}{|c|}{$\hat{m}_t\pm{\Delta_{\rm
                        stat}}^{\Delta^{2\mu_0}_{\rm
                        sys}}_{\Delta^{\mu_0/2}_{\rm sys}}\quad$
                        [GeV]}\\[2ex]
  \hline
 likelihood &  LO prediction &  NLO prediction \\ \hline
  ${\cal L}$
                      & $169.77 \pm 1.18^{+2.21}_{-2.66}$
                      & $173.65 \pm 1.20^{+0.30}_{+0.17}$
  \\ \hline
  ${\cal L}_{\rm ext}$
                      & $160.22\pm 0.34^{-6.97}_{+7.86}$
                      & $173.68\pm 0.36^{-4.11}_{+3.69}$
  \\ \hline
\end{tabular}
\caption{Extracted values for the estimator of the top-quark mass from
  9900 unweighted events following the fixed-order NLO
  prediction.\label{tab:estimators}}
\end{table}
The extracted values for the estimator of the top-quark mass together
with statistical and systematic uncertainties are summarized in
\Tab{tab:estimators}. The estimators $\hat{m}_t$ are determined from
the minima of the parabolas fitted to the negative logarithms of the
likelihood functions while the statistical uncertainties
$\Delta_{\rm stat}$ are estimated from their widths. The systematic
uncertainties $\Delta^{2\mu_0}_{\rm sys}$,
$\Delta^{\mu_0/2}_{\rm sys}$ are estimated by varying the scale by a
factor 2 around $\mu_0$. As can be seen from \Fig{fig:memunwgtev} and
\Tab{tab:estimators}, both the NLO and the LO analyses have similar
statistical uncertainties of about $1.2$ GeV and $0.35$ GeV depending
on whether the likelihood or the extended likelihood is employed. As
expected, the statistical uncertainties are to good approximation
independent from the perturbative order of the theoretical predictions
of the cross sections.  Taking the statistical uncertainties into
account, the extracted estimators from the NLO analyses are in perfect
agreement with the input value. For the likelihood as well as for the
extended likelihood the NLO differential cross section matches the
probability distribution underlying the event sample thus leading to
an unbiased estimator.  Obviously, taking into account the information
on the total number of events via the extended likelihood leads to a
reduction of the statistical uncertainties as additional information
contained in the event sample is used. Since the cross section shows a
much stronger residual scale dependence than the normalized
distributions, the extended likelihood leads however to a
significantly larger systematic uncertainty due to uncalculated higher
order corrections. In addition, the uncertainty of the luminosity
measurement which is not taken into account in the extended likelihood
analysis leads to an additional uncertainty outweighing the gain in
the reduced statistical uncertainty.

The estimators from the LO analyses on the other hand show a bias of
$2.9\times\Delta_{\rm stat}$ and $38\times\Delta_{\rm stat}$ depending
on whether the likelihood or the extended likelihood is used. It
should be emphasized that the occurrence of a bias per se does not
rule out the application of the MEM. It is well known, that the MEM
typically leads to a bias if the probability distribution used in the
evaluation of the likelihood does not match the distribution
underlying the event sample.  However, via a calibration procedure it
is possible to compensate the bias and obtain an unbiased
determination.  Since the calibration can introduce additional
uncertainties the preferred situation is that the probability
distribution used in the likelihood determination matches the
probability distribution of the event sample as best as possible thus
reducing the need of additional calibration.  As shown in
section~\ref{sec:ttbarLHC}, the NLO corrections dominantly alter the
normalization of the kinematic distributions rather than their
shape. Accordingly, the analysis employing extended likelihoods which
is sensitive to the total cross section shows thus a much stronger
separation between the results obtained from the NLO and LO
predictions.
 
Significant improvement from taking NLO corrections into account can
be seen in their impact on the theoretical uncertainties: In the NLO
analyses the theoretical uncertainties due to uncalculated higher
order corrections are roughly halved with respect to the LO analyses.
\begin{figure}
  \begin{center}
    \includegraphics[width=\columnwidth]{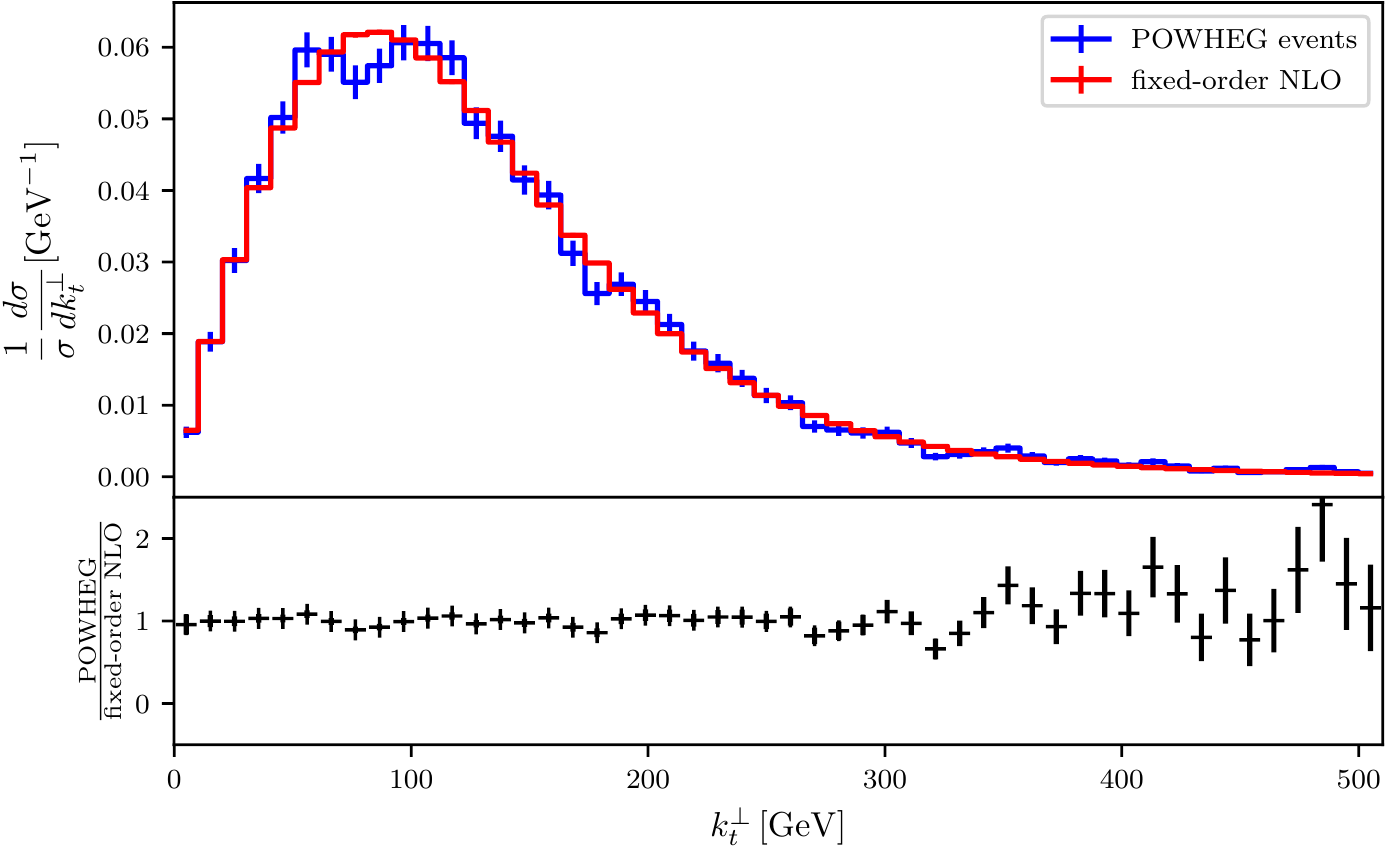}\\
    \includegraphics[width=\columnwidth]{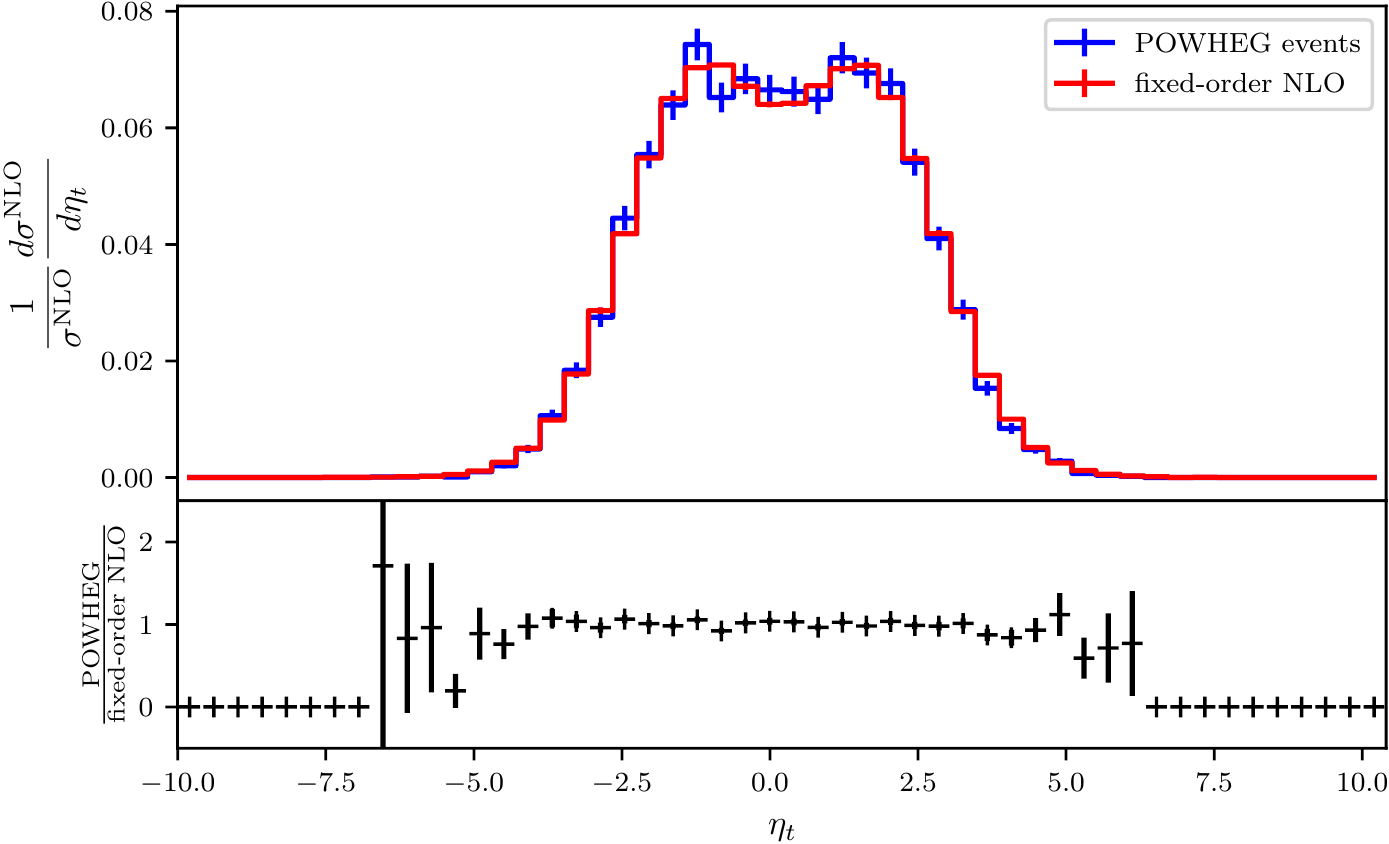}\\
    \includegraphics[width=\columnwidth]{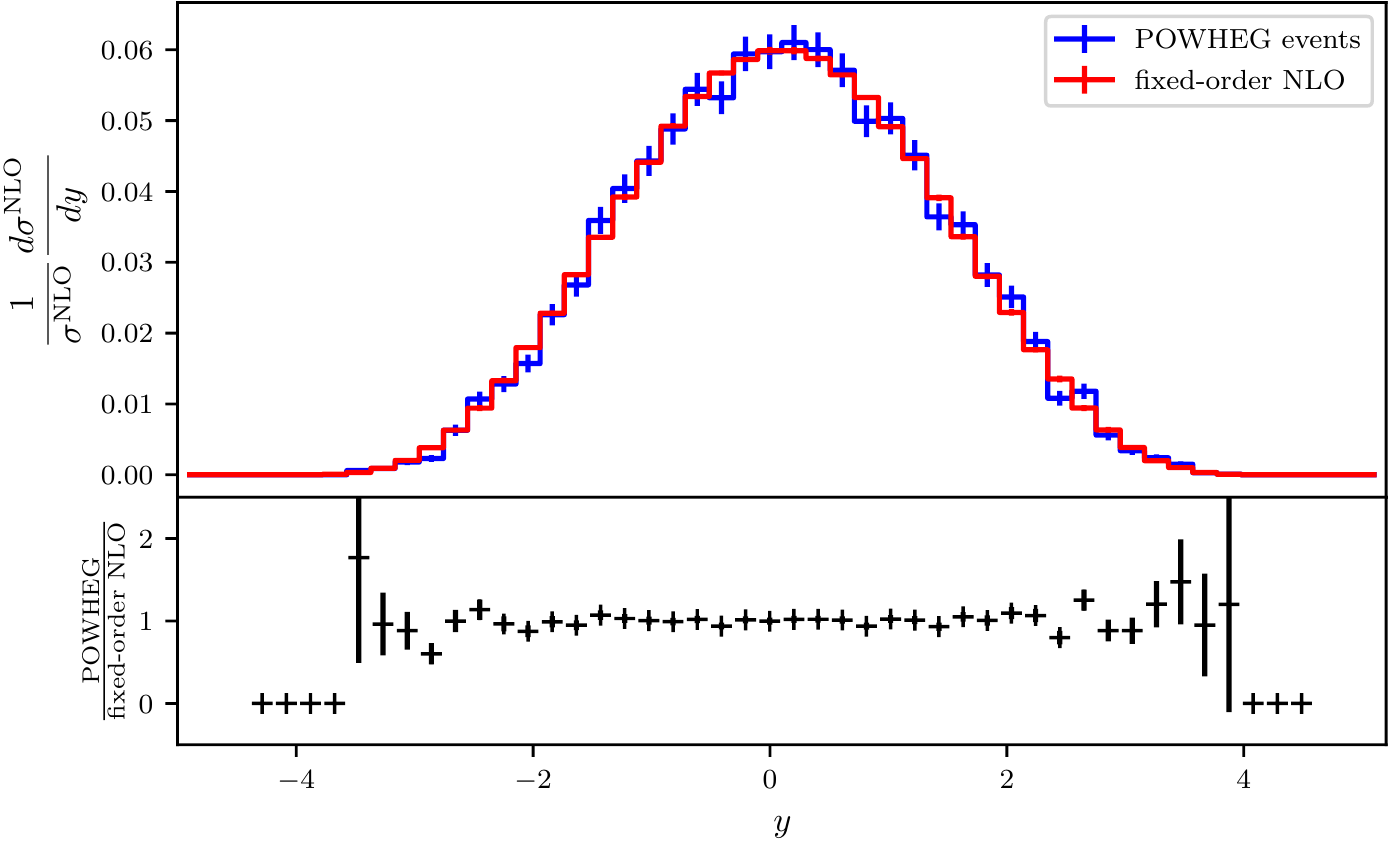}\\
  \end{center}
\caption{Impact of the parton shower on the kinematic distributions of
  the
  top quark.\label{fig:pseffects1}}
\end{figure}

In order to further study the robustness of the approach and having a
more realistic simulation, unweighted events obtained from a parton
shower simulation matched to the NLO calculation can be used. The
parton shower resums certain logarithmic corrections to all orders on
top of the fixed-order NLO parton level calculation. Since these
additional corrections present in the event sample are not accounted
for in the fixed-order-only likelihood calculation based on
\Eq{eq:evwgtNLO}, there is a mismatch between the underlying
probability distribution of the generated events and the basis of the
likelihood calculation (\Eq{eq:evwgtNLO}). As seen before in case of
the LO analysis, this mismatch can cause a systematic bias in the
extracted estimator.
 
\Fig{fig:pseffects1} shows the distributions obtained using
\texttt{POWHEG+Pythia} \cite{Nason:2004rx,Sjostrand:2006za,Frixione:2007vw,Frixione:2007nw,
  Alioli:2010xd} to generate about the same number of events as
in the case of the fixed-order analysis.  The parton shower only
mildly affects the kinematic distributions relevant for the event
definition. Further distributions supporting this observation are
shown in \Fig{fig:pseffects2} in the appendix
\ref{AdditionaCrossChecks}.  Apart from minor differences in the
$k_t^\perp$ distribution at low $k_t^\perp$, a small difference is
visible for $k_t^\perp>300$ GeV, where the \texttt{POWHEG+Pythia}
events lead to a slightly harder distribution than the events
generated from the fixed order NLO cross section.

\begin{figure}
\includegraphics[width=\columnwidth]{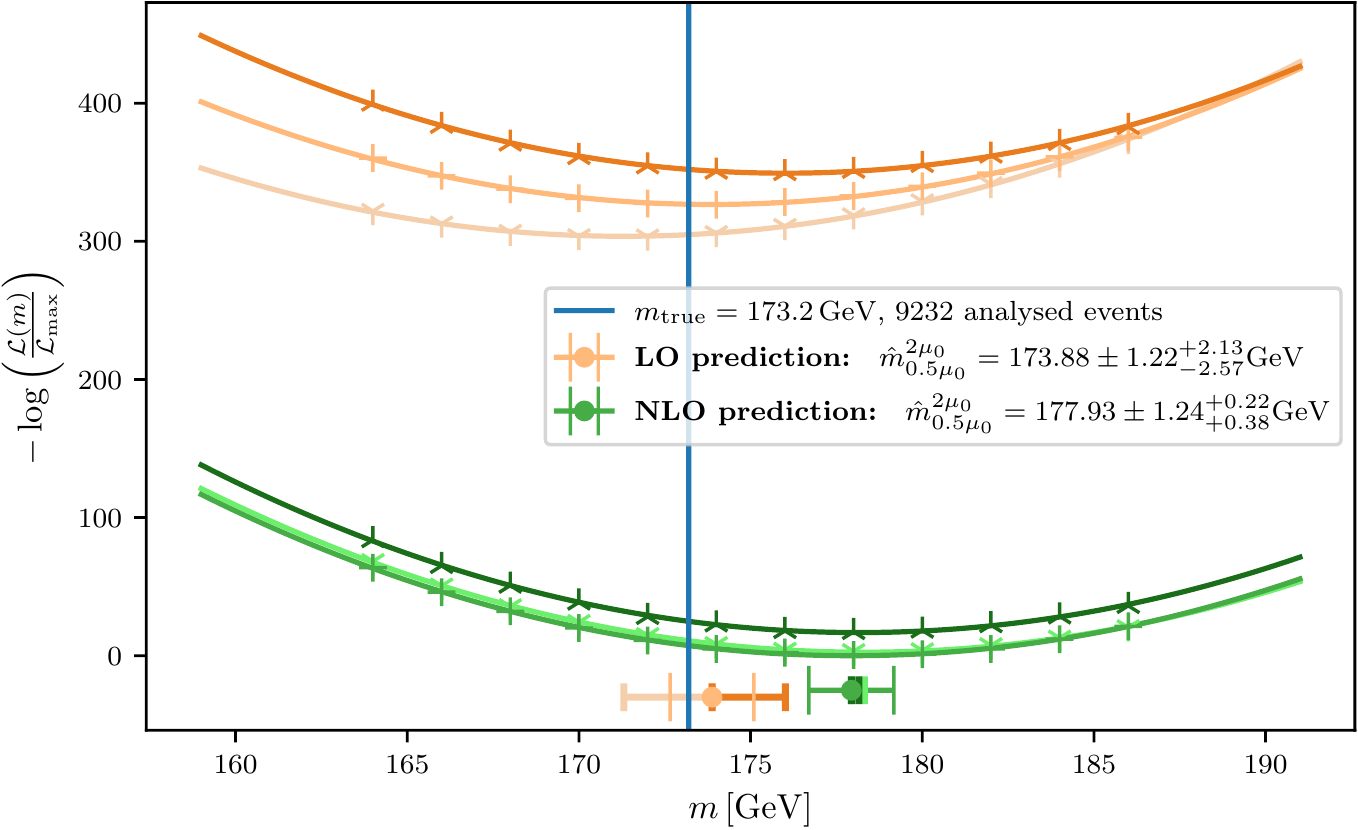}
\caption{Analysis of $9232$ \texttt{POWHEG+Pythia} events with
  fixed-order likelihoods calculated at LO and NLO
  accuracy.\label{fig:powheganalysis}}
\end{figure}

The result of the likelihood analysis using LO and NLO cross section
predictions is shown in \Fig{fig:powheganalysis} and summarized in
\Tab{tab:powhegestimators}.  We do not study the extended likelihood,
since the extended likelihood leads to much larger systematic
uncertainties.  Again the statistical uncertainties are very similar
for the LO and NLO analysis, while the systematic uncertainty is
significantly reduced when using NLO predictions.  In both cases we
observe a shift of about 4 GeV compared to the results based on the
event sample generated from the fixed-order NLO predictions. The large
shift shows the high sensitivity of the MEM with respect to tiny
changes in the distributions. In a mass determination from events
registered at the LHC this shift must be taken into account via a
calibration procedure. It is remarkable that the shift is, taking the
uncertainties into account, independent from the perturbative order of
the employed likelihood calculation.  This is similar to what has been
observed in \Refs{Martini:2017ydu, Kraus:2019qoq}. The LO likelihood
analysis reproduces the true mass value used in the
\texttt{POWHEG+Pythia} analysis. However, this ist most likely
accidental and due to the fact that the LO fixed-order results
undershoots the true mass value by about 4 GeV which is compensated by
the aforementioned shift.
\begin{table}
  \centering
\begin{tabular}{|r| r | r |}
\cline{2-3}
 \multicolumn{1}{c|}{}& \multicolumn{2}{|c|}{}\\[-2ex]
 \multicolumn{1}{c|}{}& \multicolumn{2}{|c|}{$\hat{m}_t\pm{\Delta_{\rm stat}}^{\Delta^{2\mu_0}_{\rm sys}}_{\Delta^{\mu_0/2}_{\rm sys}}\quad$  [GeV]}\\[2ex] \hline
 likelihood &  LO prediction &  NLO prediction \\ \hline
${\cal L}$ 
	& $173.88 \pm 1.22^{+2.13}_{-2.57}$  
	&  $177.93 \pm 1.24^{+0.22}_{+0.38}$ \\ \hline
\end{tabular}
\caption{Extracted values for the estimator of the top-quark mass from
  unweighted \texttt{POWHEG+Pythia} events following the NLO prediction
  matched to a parton shower. \label{tab:powhegestimators}}
\end{table}

\section{Conclusion\label{sec:concl}}
In this work the MEM at NLO is applied to top-quark pair production at
the LHC. To investigate the potential of the matrix element method to
measure the top-quark mass, the MEM at NLO is applied to pseudo-data:
unweighted events generated from the fixed-order NLO cross section as
well as events obtained using \texttt{POWHEG+Pythia} incorporating the
parton shower effects.  Using pseudo-data based on
\texttt{POWHEG+Pythia} allows to study the effect of the parton shower
and gives a more realistic simulation. Including the NLO corrections
in the likelihood calculation leads to a significant reduction of the
theoretical uncertainties of the extracted top-quark mass, while the
statistical uncertainties remain almost unchanged compared to the LO
analysis.  We stress that the uncertainties due to scale variation
cannot be reduced by a calibration. The reduction of the uncertainties
associated with the scale variation when going from LO to NLO thus
presents an important improvement and a strong argument in favour of
the the MEM at NLO accuracy.

Another important observation is the fact that the extended likelihood
yields a significant improvement in terms of the statistical
uncertainties. However, in practical applications this gain in
precision is completely outweighed by the theoretical uncertainties of
the number of expected events.  This can be understood from the fact
that, much as the NLO corrections (see \Fig{fig:kfacs1}), the scale
variations do not dramatically change the shape of the kinematic
distributions but mostly their normalization (see \Fig{fig:scalevar})
thereby making the extended likelihood analyses more sensitive to
their effect. Additionally, employing the extended likelihood requires
precise knowledge of the integrated luminosity of the LHC. The
dependence on this parameter introduces an additional source of
systematic uncertainty. This has to be taken into account for future
experimental applications of the MEM with realistic event numbers for
abundantly produced top-quark pairs at the LHC which will most likely
be dominated by systematic uncertainties.  As has already been stated
before
(\cite{Martini:2015fsa,Martini:2017ydu,Kraus:2019qoq,Kraus:2019myc}),
for parameter inference with the MEM it is mandatory to perform the
likelihood calculation at least at NLO accuracy in order to properly
fix the renormalization scheme of the extracted parameter.

The application of the MEM at NLO to top-quark pair events at the LHC
can offer an alternative approach to determine the top-quark mass with
high accuracy. As has been demonstrated in this work, already for 
a few ten thousand events the precision of the analysis becomes
dominated by systematic uncertainties. As the LHC produces millions of
top-quark pairs, the analysis could be performed with a rather small
fraction of cherry-picked events allowing to minimize the overall
systematic uncertainty. The results obtained in this article suggest
that top-quark mass determination with an uncertainty below 1 GeV
could be feasible.  Of course, for a realistic application of the MEM
to experimental data, transfer functions accounting for decays,
additional radiation and detector effects have to be considered. In
addition, as the analysis based on the events including parton shower
effects shows, a further calibration is required.

\section*{Acknowledgments}
This work was supported in part by the German Federal
Ministry of Education and Research (Bundesministerium für
Bildung und Forschung) under Contract No. 05H18KHCA1

\appendix
\section{Additional results on distributions used for the
  validation\label{AdditionaCrossChecks}}

\begin{figure}
  \begin{center}
    \includegraphics[width=\columnwidth]{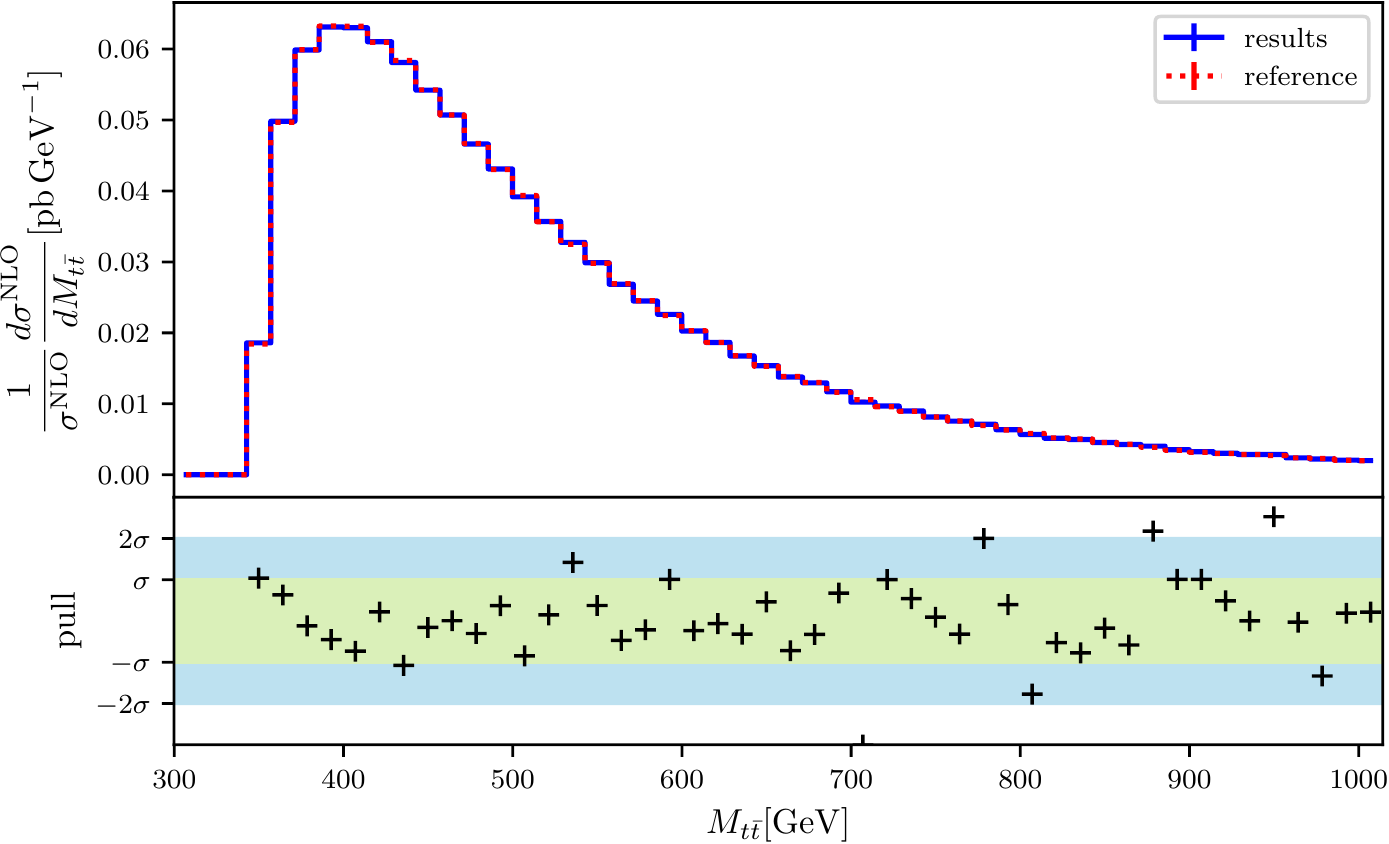}\\
    \includegraphics[width=\columnwidth]{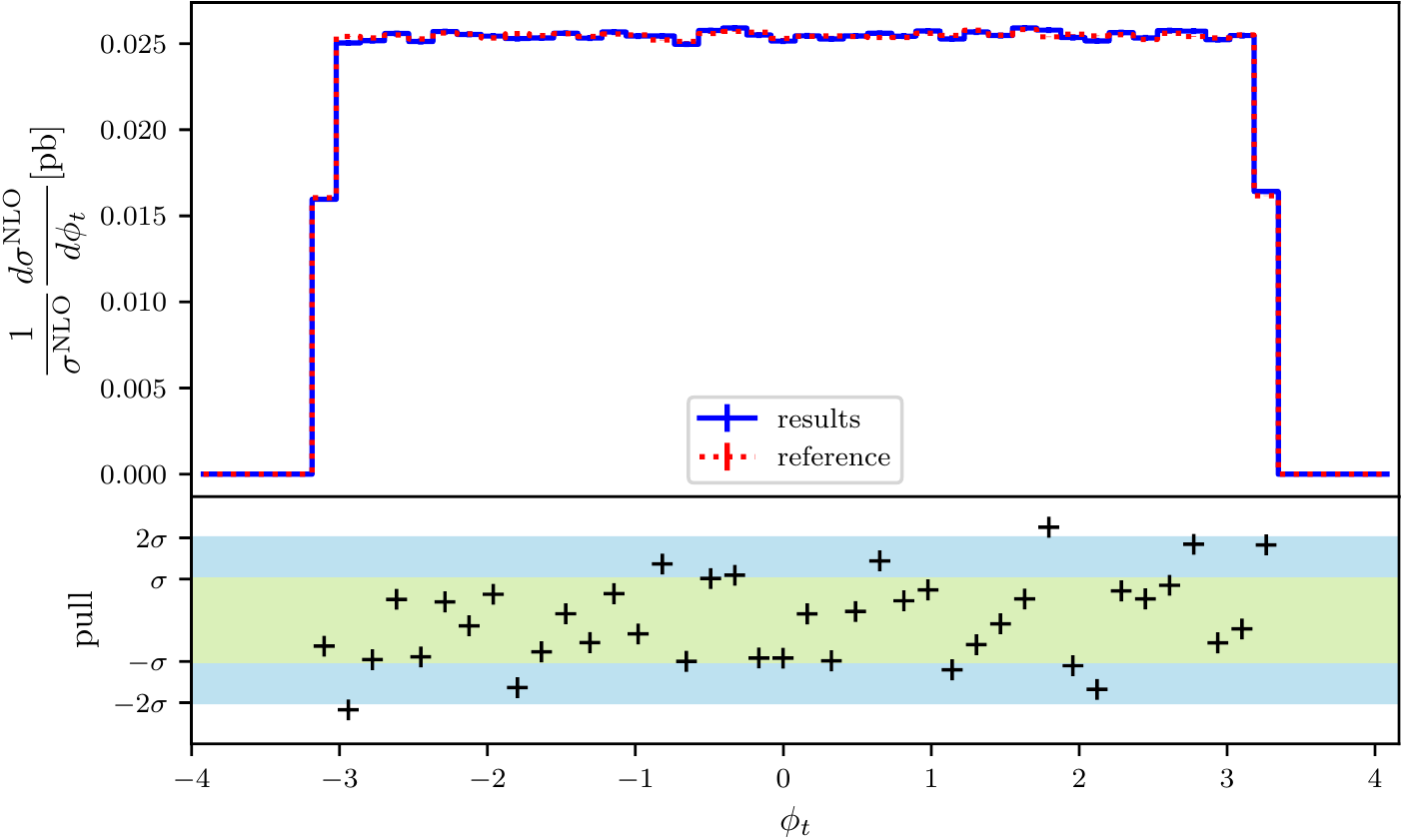}
  \end{center}
  \caption{Same as \Fig{fig:evvalid1} but for the  $M_{t\bar t}$- and
    the $\phi_t$-distribution.\label{fig:evvalid2}}
\end{figure}

\begin{figure}
  \begin{center}
    \includegraphics[width=\columnwidth]{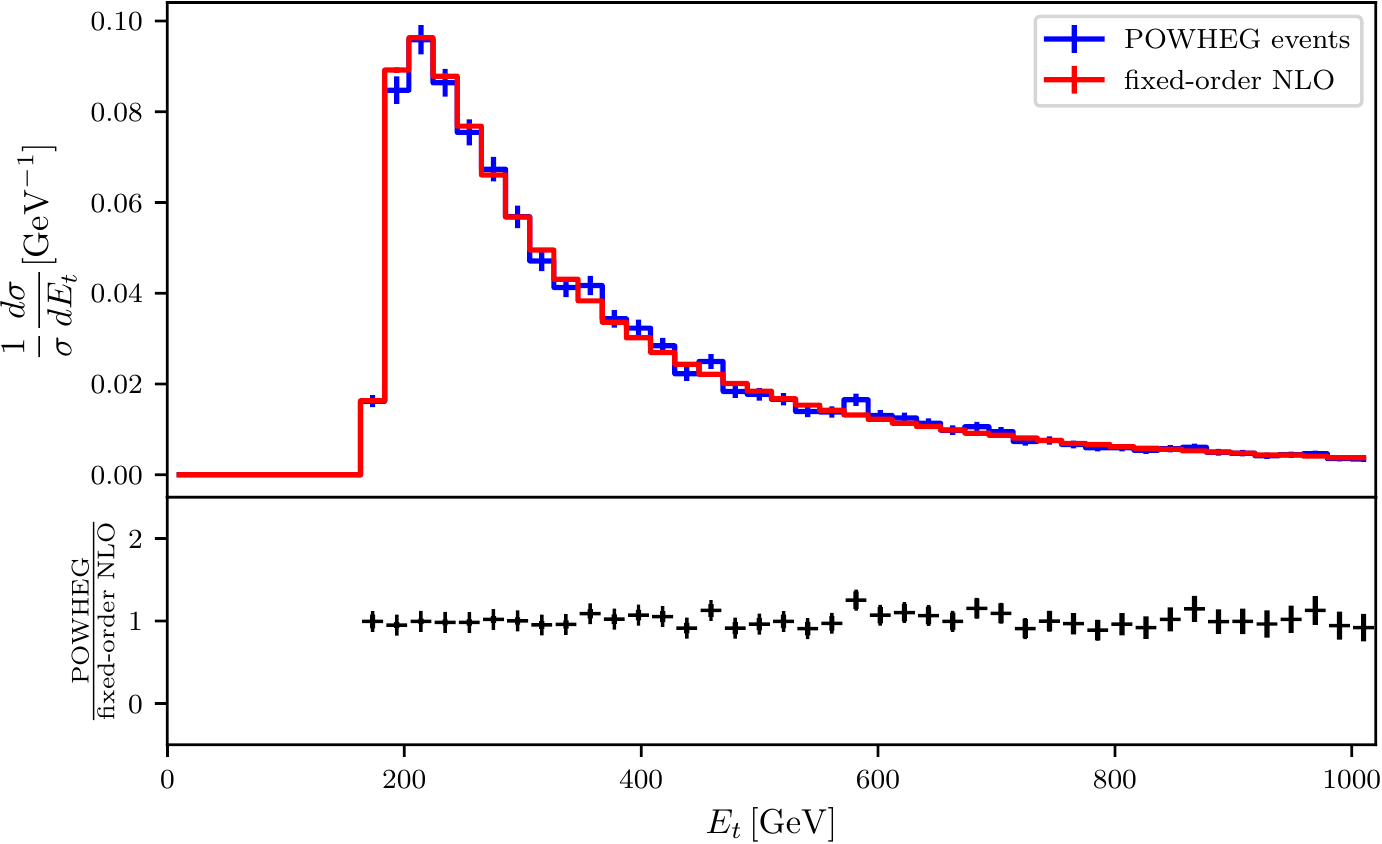}\\
    \includegraphics[width=\columnwidth]{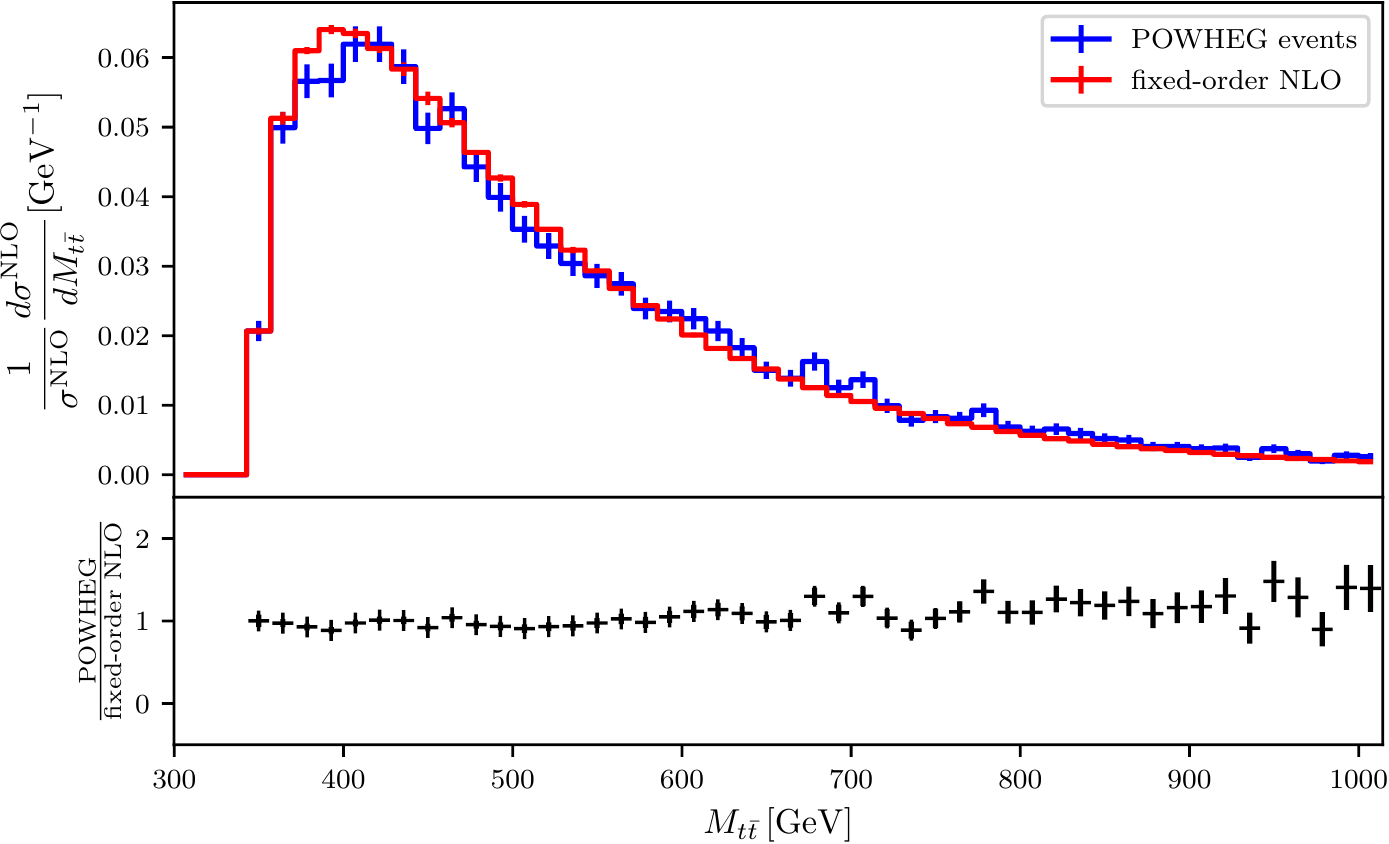}\\
    \includegraphics[width=\columnwidth]{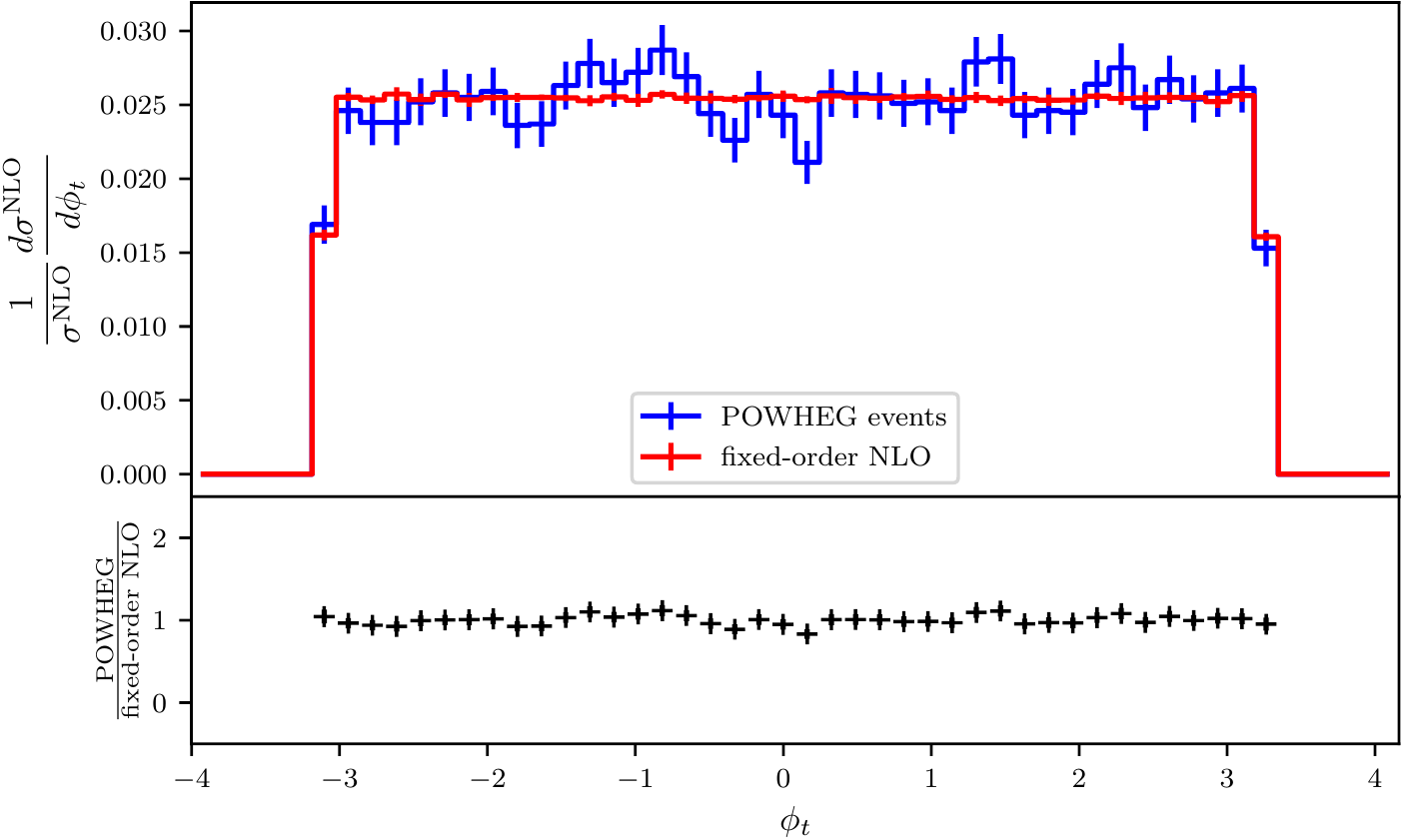}\\
  \end{center}
  \caption{Same as \Fig{fig:pseffects1} but for the energy, $M_{t\bar t}$- and
    $\phi_t$-distribution.\label{fig:pseffects2}}
\end{figure}

In this appendix we show further cross checks used for the validation
of the implementation. 
\begin{figure}
  \begin{center}
  \includegraphics[width=\columnwidth]{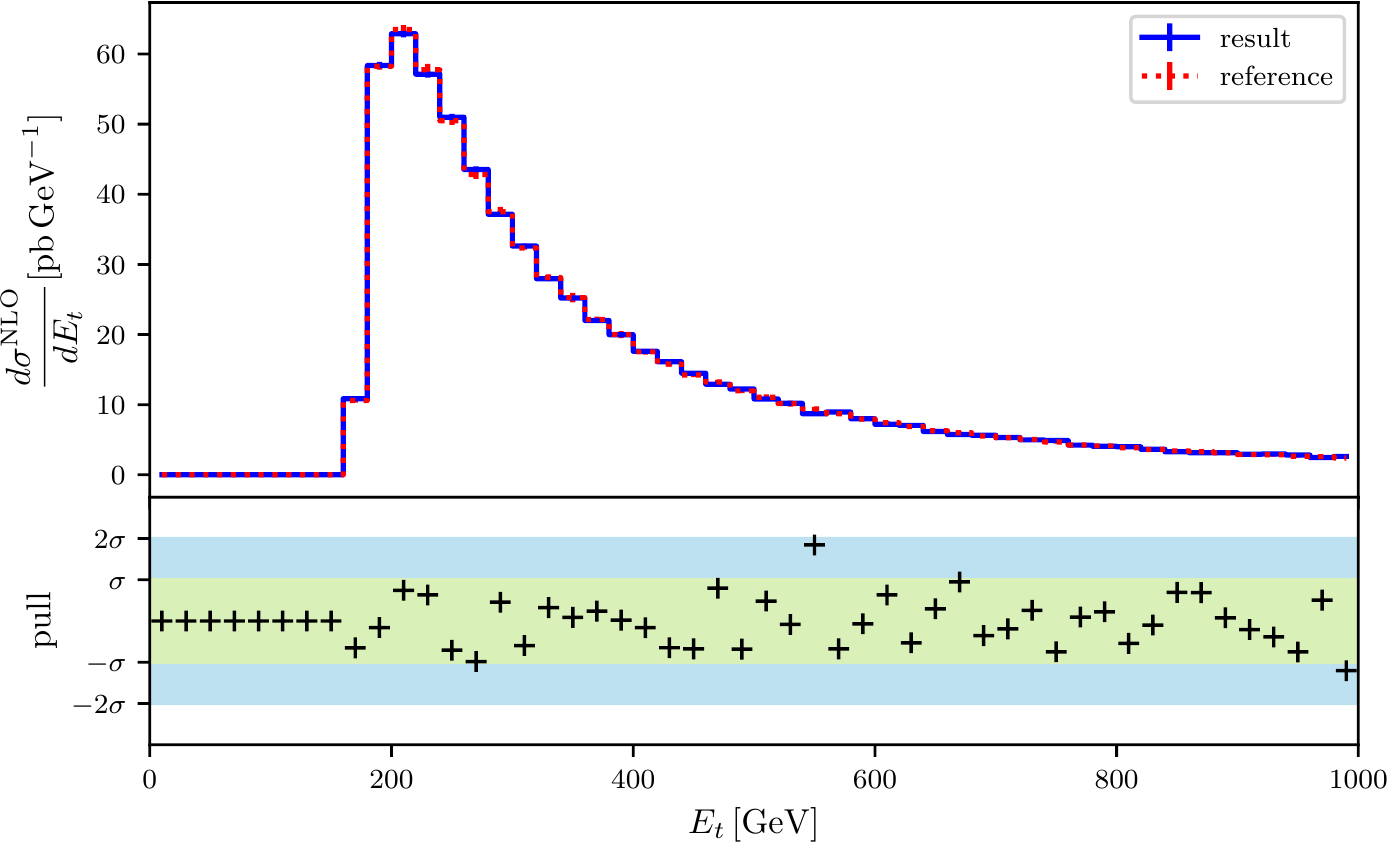}\\
  \includegraphics[width=\columnwidth]{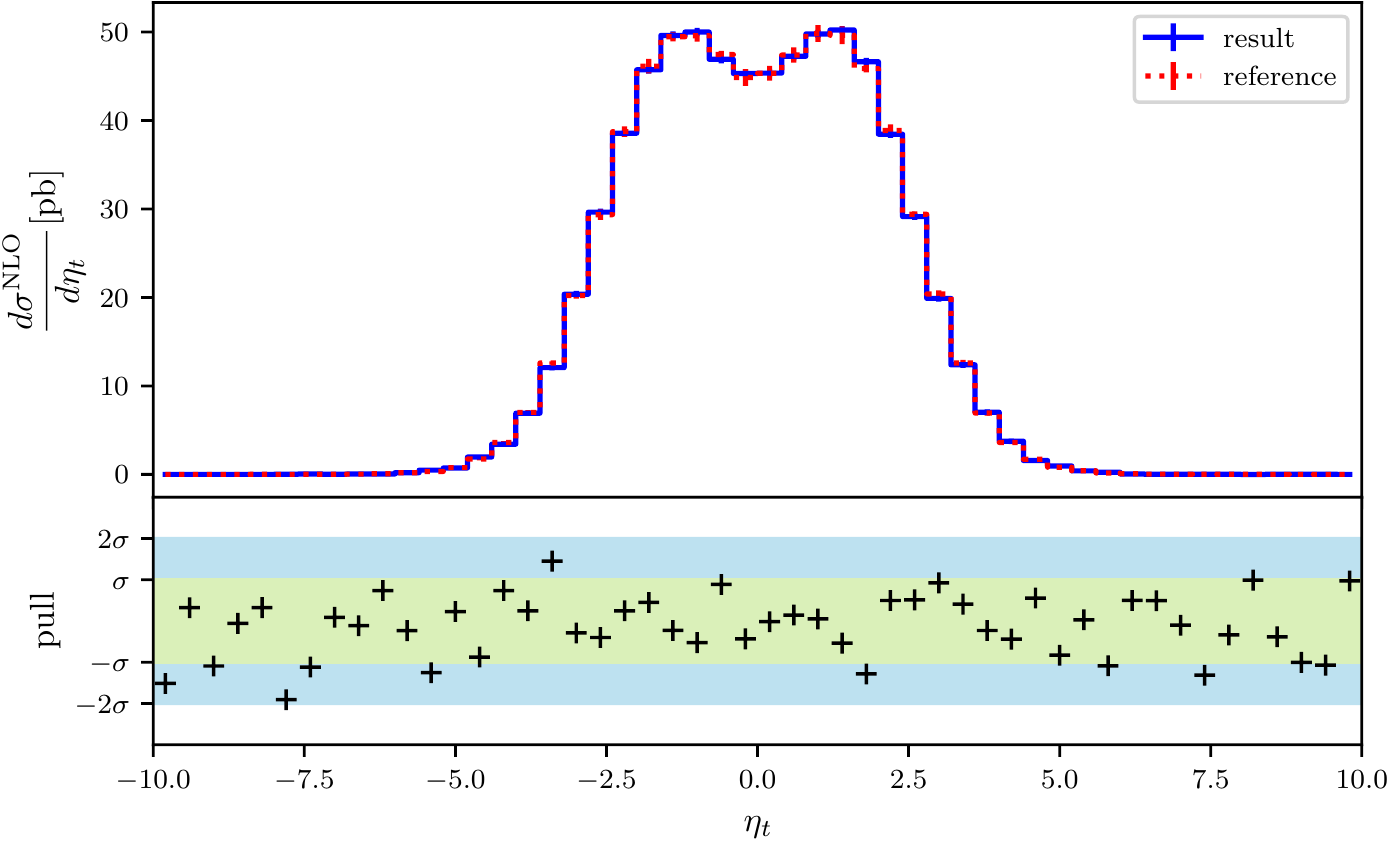}    
  \end{center}
\caption{Validation of the implementation: Comparison of differential
  distributions of the top quark obtained in this work with results
  from \texttt{madgraph5 aMC@NLO}.\label{fig:diffvalid1}}
\end{figure}
\begin{figure}
  \begin{center}
  \includegraphics[width=\columnwidth]{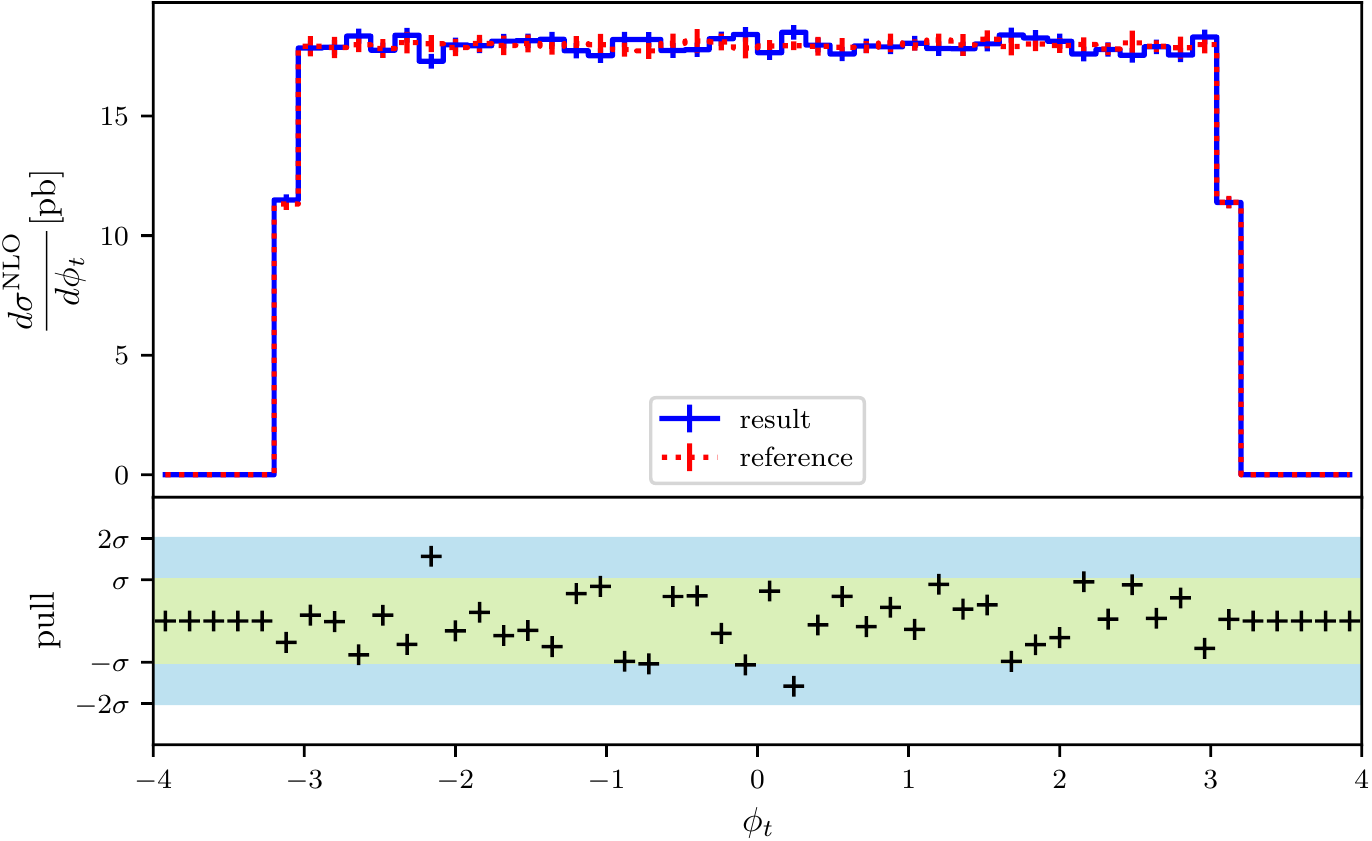}\\
  \includegraphics[width=\columnwidth]{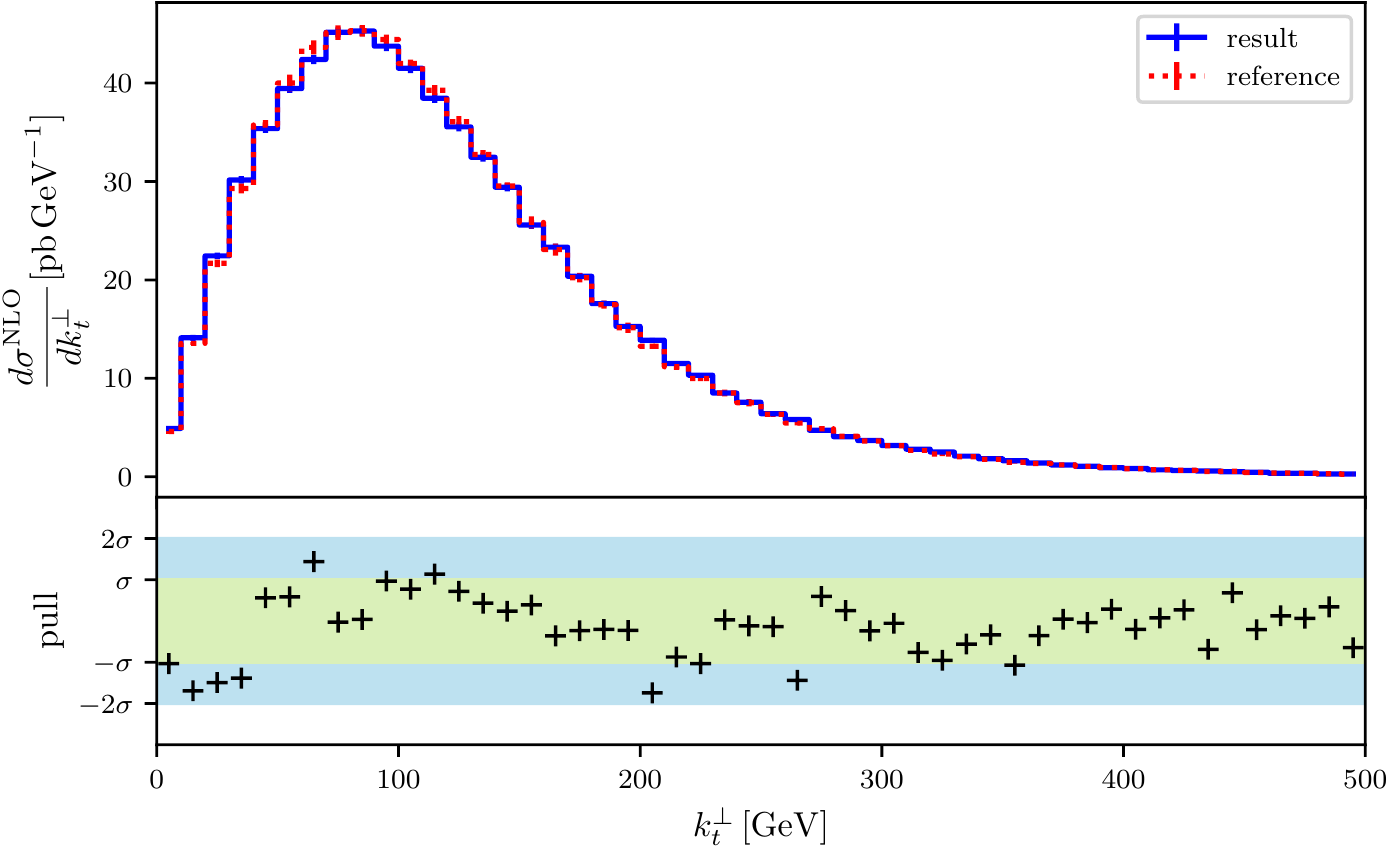}    
  \end{center}
  \caption{Same as \Fig{fig:diffvalid1} but for the $\phi_t$- and
    the $k_t^\perp$-distribution.\label{fig:diffvalid2}}
\end{figure}
\Fig{fig:diffvalid1} shows comparisons of NLO predictions for
differential distributions calculated in this work with distributions
obtained from \texttt{madgraph5 aMC@NLO} \cite{Alwall:2014hca} which is
based on the dipole subtraction method
\cite{Catani:1996vz,Catani:2002hc}. The pull distributions in the
bottom plots of \Fig{fig:diffvalid1} and \Fig{fig:diffvalid2}
illustrate the agreement between both implementations within
statistical uncertainties. This comparisons serve as a further
validation for the choice of the slicing parameter.
\begin{figure}
  \begin{center}
    \includegraphics[width=\columnwidth]{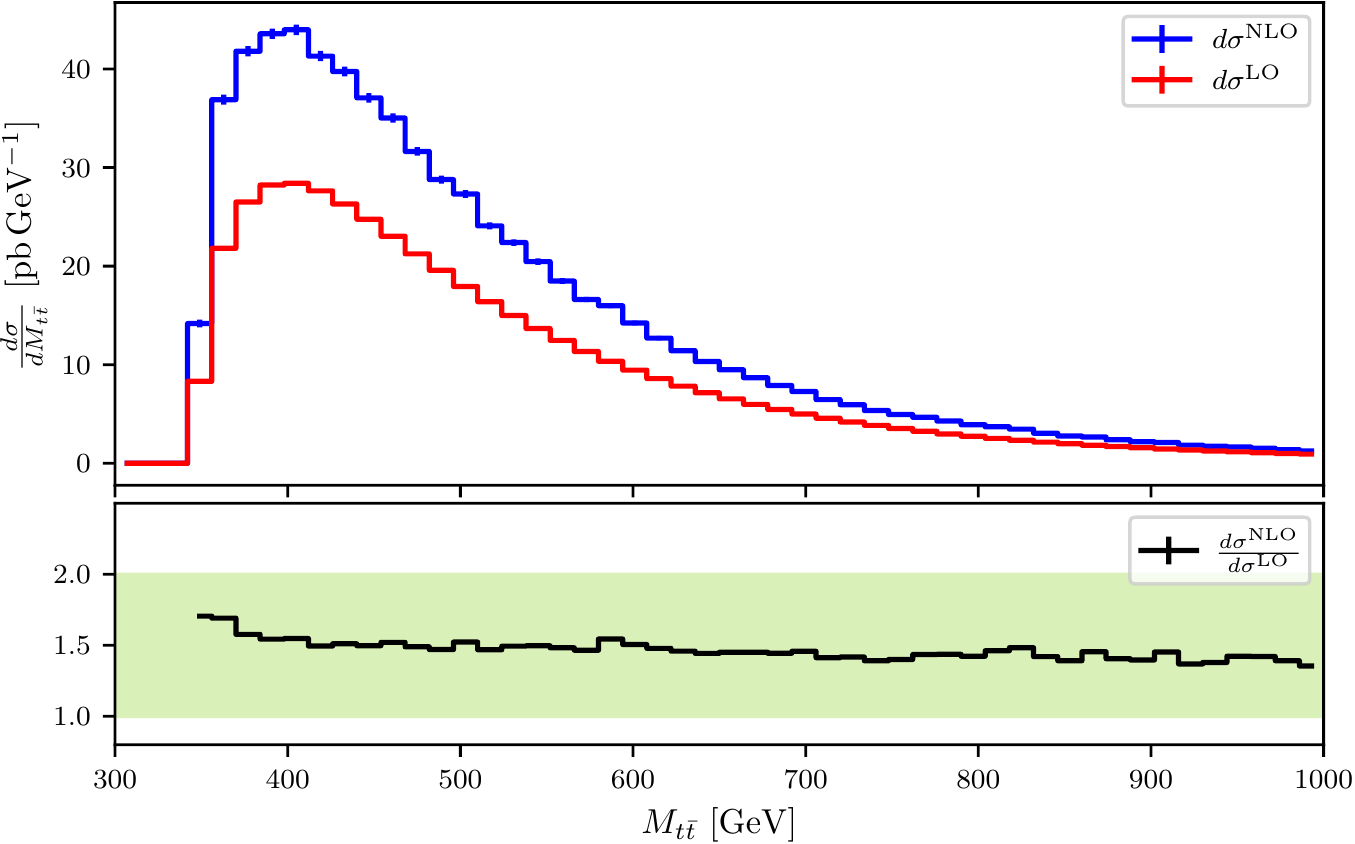}\\
    \includegraphics[width=\columnwidth]{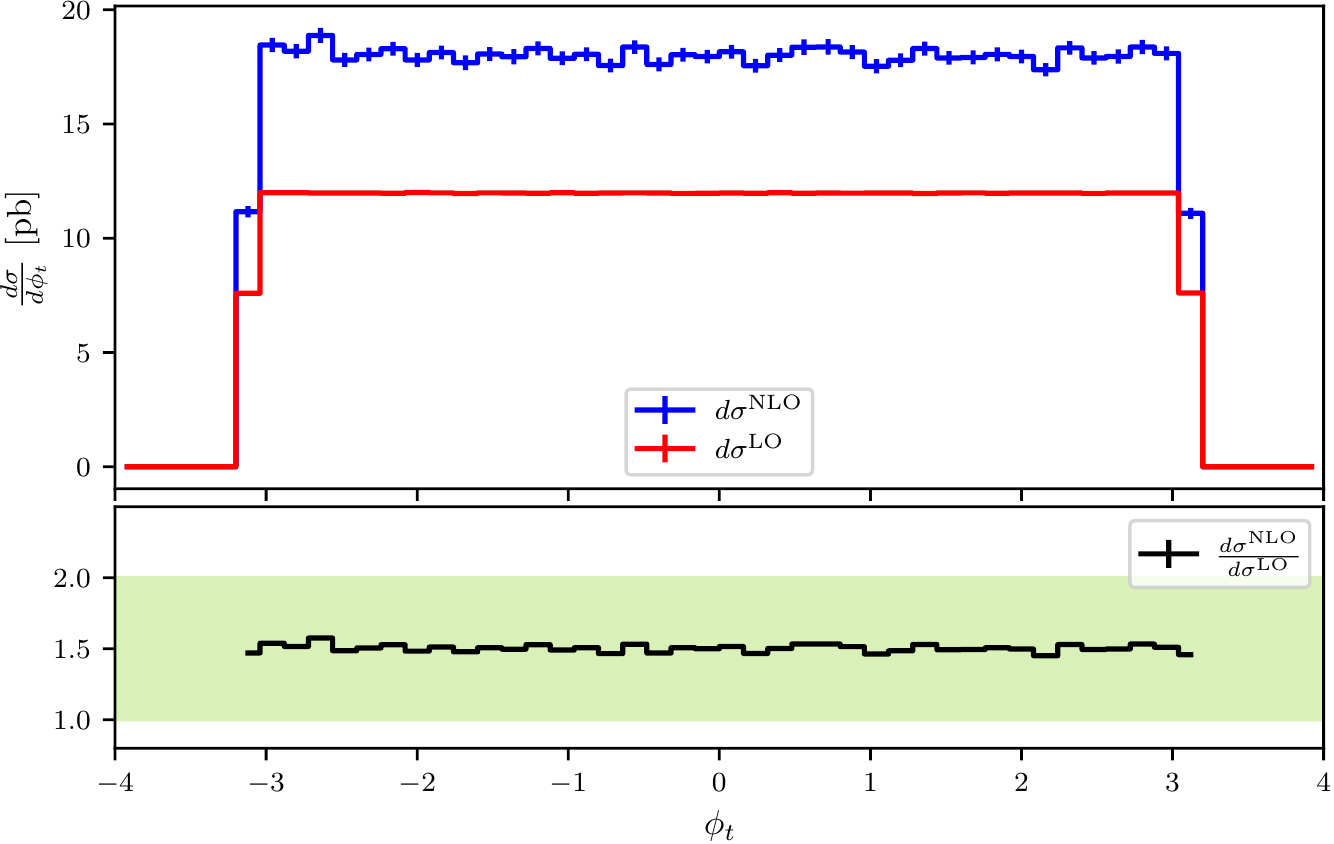}
  \end{center}
  \caption{Same as \Fig{fig:kfacs1} but for the $M_{t\bar t}$ and
    the $\phi_t$-distribution.\label{fig:kfacs2}}
\end{figure}
\Fig{fig:kfacs2} shows the NLO corrections (upper part) together with
the k-factors (lower part) for the $M_{t\bar t}$ and the
$\phi_t$-distribution. Similar to what is shown in \Fig{fig:kfacs1}
again a flat k-factor is observed. As a check of the event generation
and the unweighting procedure \Fig{fig:evvalid2} shows distributions
calculated from the generated unweighted events compared with a
calculation using \texttt{madgraph5 aMC@NLO} \cite{Alwall:2014hca}.
Similar to \Fig{fig:pseffects1} we show in \Fig{fig:pseffects2} for
further distributions the comparison of distributions obtained at
fixed-order NLO accuracy with results using \texttt{POWHEG+Pythia}.
\clearpage
\bibliographystyle{apsrev4-1}
\bibliography{lit}

\begin{thebibliography}{43}%
\makeatletter
\providecommand \@ifxundefined [1]{%
 \@ifx{#1\undefined}
}%
\providecommand \@ifnum [1]{%
 \ifnum #1\expandafter \@firstoftwo
 \else \expandafter \@secondoftwo
 \fi
}%
\providecommand \@ifx [1]{%
 \ifx #1\expandafter \@firstoftwo
 \else \expandafter \@secondoftwo
 \fi
}%
\providecommand \natexlab [1]{#1}%
\providecommand \enquote  [1]{``#1''}%
\providecommand \bibnamefont  [1]{#1}%
\providecommand \bibfnamefont [1]{#1}%
\providecommand \citenamefont [1]{#1}%
\providecommand \href@noop [0]{\@secondoftwo}%
\providecommand \href [0]{\begingroup \@sanitize@url \@href}%
\providecommand \@href[1]{\@@startlink{#1}\@@href}%
\providecommand \@@href[1]{\endgroup#1\@@endlink}%
\providecommand \@sanitize@url [0]{\catcode `\\12\catcode `\$12\catcode
  `\&12\catcode `\#12\catcode `\^12\catcode `\_12\catcode `\%12\relax}%
\providecommand \@@startlink[1]{}%
\providecommand \@@endlink[0]{}%
\providecommand \url  [0]{\begingroup\@sanitize@url \@url }%
\providecommand \@url [1]{\endgroup\@href {#1}{\urlprefix }}%
\providecommand \urlprefix  [0]{URL }%
\providecommand \Eprint [0]{\href }%
\providecommand \doibase [0]{http://dx.doi.org/}%
\providecommand \selectlanguage [0]{\@gobble}%
\providecommand \bibinfo  [0]{\@secondoftwo}%
\providecommand \bibfield  [0]{\@secondoftwo}%
\providecommand \translation [1]{[#1]}%
\providecommand \BibitemOpen [0]{}%
\providecommand \bibitemStop [0]{}%
\providecommand \bibitemNoStop [0]{.\EOS\space}%
\providecommand \EOS [0]{\spacefactor3000\relax}%
\providecommand \BibitemShut  [1]{\csname bibitem#1\endcsname}%
\let\auto@bib@innerbib\@empty
\bibitem [{\citenamefont {Nason}\ \emph {et~al.}(1988)\citenamefont {Nason},
  \citenamefont {Dawson},\ and\ \citenamefont {Ellis}}]{Nason:1987xz}%
  \BibitemOpen
  \bibfield  {author} {\bibinfo {author} {\bibfnamefont {P.}~\bibnamefont
  {Nason}}, \bibinfo {author} {\bibfnamefont {S.}~\bibnamefont {Dawson}}, \
  and\ \bibinfo {author} {\bibfnamefont {R.~K.}\ \bibnamefont {Ellis}},\ }\href
  {\doibase 10.1016/0550-3213(88)90422-1} {\bibfield  {journal} {\bibinfo
  {journal} {Nucl. Phys. B}\ }\textbf {\bibinfo {volume} {303}},\ \bibinfo
  {pages} {607} (\bibinfo {year} {1988})}\BibitemShut {NoStop}%
\bibitem [{\citenamefont {Nason}\ \emph {et~al.}(1989)\citenamefont {Nason},
  \citenamefont {Dawson},\ and\ \citenamefont {Ellis}}]{Nason:1989zy}%
  \BibitemOpen
  \bibfield  {author} {\bibinfo {author} {\bibfnamefont {P.}~\bibnamefont
  {Nason}}, \bibinfo {author} {\bibfnamefont {S.}~\bibnamefont {Dawson}}, \
  and\ \bibinfo {author} {\bibfnamefont {R.~K.}\ \bibnamefont {Ellis}},\ }\href
  {\doibase 10.1016/0550-3213(89)90286-1} {\bibfield  {journal} {\bibinfo
  {journal} {Nucl. Phys. B}\ }\textbf {\bibinfo {volume} {327}},\ \bibinfo
  {pages} {49} (\bibinfo {year} {1989})},\ \bibinfo {note} {[Erratum:
  Nucl.Phys.B 335, 260--260 (1990)]}\BibitemShut {NoStop}%
\bibitem [{\citenamefont {Beenakker}\ \emph {et~al.}(1989)\citenamefont
  {Beenakker}, \citenamefont {Kuijf}, \citenamefont {van Neerven},\ and\
  \citenamefont {Smith}}]{Beenakker:1988bq}%
  \BibitemOpen
  \bibfield  {author} {\bibinfo {author} {\bibfnamefont {W.}~\bibnamefont
  {Beenakker}}, \bibinfo {author} {\bibfnamefont {H.}~\bibnamefont {Kuijf}},
  \bibinfo {author} {\bibfnamefont {W.~L.}\ \bibnamefont {van Neerven}}, \ and\
  \bibinfo {author} {\bibfnamefont {J.}~\bibnamefont {Smith}},\ }\href
  {\doibase 10.1103/PhysRevD.40.54} {\bibfield  {journal} {\bibinfo  {journal}
  {Phys. Rev. D}\ }\textbf {\bibinfo {volume} {40}},\ \bibinfo {pages} {54}
  (\bibinfo {year} {1989})}\BibitemShut {NoStop}%
\bibitem [{\citenamefont {Beenakker}\ \emph {et~al.}(1991)\citenamefont
  {Beenakker}, \citenamefont {van Neerven}, \citenamefont {Meng}, \citenamefont
  {Schuler},\ and\ \citenamefont {Smith}}]{Beenakker:1990maa}%
  \BibitemOpen
  \bibfield  {author} {\bibinfo {author} {\bibfnamefont {W.}~\bibnamefont
  {Beenakker}}, \bibinfo {author} {\bibfnamefont {W.~L.}\ \bibnamefont {van
  Neerven}}, \bibinfo {author} {\bibfnamefont {R.}~\bibnamefont {Meng}},
  \bibinfo {author} {\bibfnamefont {G.~A.}\ \bibnamefont {Schuler}}, \ and\
  \bibinfo {author} {\bibfnamefont {J.}~\bibnamefont {Smith}},\ }\href
  {\doibase 10.1016/S0550-3213(05)80032-X} {\bibfield  {journal} {\bibinfo
  {journal} {Nucl. Phys. B}\ }\textbf {\bibinfo {volume} {351}},\ \bibinfo
  {pages} {507} (\bibinfo {year} {1991})}\BibitemShut {NoStop}%
\bibitem [{\citenamefont {Bernreuther}\ \emph {et~al.}(2004)\citenamefont
  {Bernreuther}, \citenamefont {Brandenburg}, \citenamefont {Si},\ and\
  \citenamefont {Uwer}}]{Bernreuther:2004jv}%
  \BibitemOpen
  \bibfield  {author} {\bibinfo {author} {\bibfnamefont {W.}~\bibnamefont
  {Bernreuther}}, \bibinfo {author} {\bibfnamefont {A.}~\bibnamefont
  {Brandenburg}}, \bibinfo {author} {\bibfnamefont {Z.~G.}\ \bibnamefont {Si}},
  \ and\ \bibinfo {author} {\bibfnamefont {P.}~\bibnamefont {Uwer}},\ }\href
  {\doibase 10.1016/j.nuclphysb.2004.04.019} {\bibfield  {journal} {\bibinfo
  {journal} {Nucl. Phys. B}\ }\textbf {\bibinfo {volume} {690}},\ \bibinfo
  {pages} {81} (\bibinfo {year} {2004})},\ \Eprint
  {http://arxiv.org/abs/hep-ph/0403035} {arXiv:hep-ph/0403035} \BibitemShut
  {NoStop}%
\bibitem [{\citenamefont {Melnikov}\ and\ \citenamefont
  {Schulze}(2009)}]{Melnikov:2009dn}%
  \BibitemOpen
  \bibfield  {author} {\bibinfo {author} {\bibfnamefont {K.}~\bibnamefont
  {Melnikov}}\ and\ \bibinfo {author} {\bibfnamefont {M.}~\bibnamefont
  {Schulze}},\ }\href {\doibase 10.1088/1126-6708/2009/08/049} {\bibfield
  {journal} {\bibinfo  {journal} {JHEP}\ }\textbf {\bibinfo {volume} {08}},\
  \bibinfo {pages} {049} (\bibinfo {year} {2009})},\ \Eprint
  {http://arxiv.org/abs/0907.3090} {arXiv:0907.3090 [hep-ph]} \BibitemShut
  {NoStop}%
\bibitem [{\citenamefont {Czakon}\ \emph {et~al.}(2013)\citenamefont {Czakon},
  \citenamefont {Fiedler},\ and\ \citenamefont {Mitov}}]{Czakon:2013goa}%
  \BibitemOpen
  \bibfield  {author} {\bibinfo {author} {\bibfnamefont {M.}~\bibnamefont
  {Czakon}}, \bibinfo {author} {\bibfnamefont {P.}~\bibnamefont {Fiedler}}, \
  and\ \bibinfo {author} {\bibfnamefont {A.}~\bibnamefont {Mitov}},\ }\href
  {\doibase 10.1103/PhysRevLett.110.252004} {\bibfield  {journal} {\bibinfo
  {journal} {Phys. Rev. Lett.}\ }\textbf {\bibinfo {volume} {110}},\ \bibinfo
  {pages} {252004} (\bibinfo {year} {2013})},\ \Eprint
  {http://arxiv.org/abs/1303.6254} {arXiv:1303.6254 [hep-ph]} \BibitemShut
  {NoStop}%
\bibitem [{\citenamefont {Czakon}\ \emph
  {et~al.}(2016{\natexlab{a}})\citenamefont {Czakon}, \citenamefont {Heymes},\
  and\ \citenamefont {Mitov}}]{Czakon:2015owf}%
  \BibitemOpen
  \bibfield  {author} {\bibinfo {author} {\bibfnamefont {M.}~\bibnamefont
  {Czakon}}, \bibinfo {author} {\bibfnamefont {D.}~\bibnamefont {Heymes}}, \
  and\ \bibinfo {author} {\bibfnamefont {A.}~\bibnamefont {Mitov}},\ }\href
  {\doibase 10.1103/PhysRevLett.116.082003} {\bibfield  {journal} {\bibinfo
  {journal} {Phys. Rev. Lett.}\ }\textbf {\bibinfo {volume} {116}},\ \bibinfo
  {pages} {082003} (\bibinfo {year} {2016}{\natexlab{a}})},\ \Eprint
  {http://arxiv.org/abs/1511.00549} {arXiv:1511.00549 [hep-ph]} \BibitemShut
  {NoStop}%
\bibitem [{\citenamefont {Czakon}\ \emph
  {et~al.}(2016{\natexlab{b}})\citenamefont {Czakon}, \citenamefont {Fiedler},
  \citenamefont {Heymes},\ and\ \citenamefont {Mitov}}]{Czakon:2016ckf}%
  \BibitemOpen
  \bibfield  {author} {\bibinfo {author} {\bibfnamefont {M.}~\bibnamefont
  {Czakon}}, \bibinfo {author} {\bibfnamefont {P.}~\bibnamefont {Fiedler}},
  \bibinfo {author} {\bibfnamefont {D.}~\bibnamefont {Heymes}}, \ and\ \bibinfo
  {author} {\bibfnamefont {A.}~\bibnamefont {Mitov}},\ }\href {\doibase
  10.1007/JHEP05(2016)034} {\bibfield  {journal} {\bibinfo  {journal} {JHEP}\
  }\textbf {\bibinfo {volume} {05}},\ \bibinfo {pages} {034} (\bibinfo {year}
  {2016}{\natexlab{b}})},\ \Eprint {http://arxiv.org/abs/1601.05375}
  {arXiv:1601.05375 [hep-ph]} \BibitemShut {NoStop}%
\bibitem [{\citenamefont {Catani}\ \emph
  {et~al.}(2019{\natexlab{a}})\citenamefont {Catani}, \citenamefont {Devoto},
  \citenamefont {Grazzini}, \citenamefont {Kallweit}, \citenamefont
  {Mazzitelli},\ and\ \citenamefont {Sargsyan}}]{Catani:2019iny}%
  \BibitemOpen
  \bibfield  {author} {\bibinfo {author} {\bibfnamefont {S.}~\bibnamefont
  {Catani}}, \bibinfo {author} {\bibfnamefont {S.}~\bibnamefont {Devoto}},
  \bibinfo {author} {\bibfnamefont {M.}~\bibnamefont {Grazzini}}, \bibinfo
  {author} {\bibfnamefont {S.}~\bibnamefont {Kallweit}}, \bibinfo {author}
  {\bibfnamefont {J.}~\bibnamefont {Mazzitelli}}, \ and\ \bibinfo {author}
  {\bibfnamefont {H.}~\bibnamefont {Sargsyan}},\ }\href {\doibase
  10.1103/PhysRevD.99.051501} {\bibfield  {journal} {\bibinfo  {journal} {Phys.
  Rev. D}\ }\textbf {\bibinfo {volume} {99}},\ \bibinfo {pages} {051501}
  (\bibinfo {year} {2019}{\natexlab{a}})},\ \Eprint
  {http://arxiv.org/abs/1901.04005} {arXiv:1901.04005 [hep-ph]} \BibitemShut
  {NoStop}%
\bibitem [{\citenamefont {Catani}\ \emph
  {et~al.}(2019{\natexlab{b}})\citenamefont {Catani}, \citenamefont {Devoto},
  \citenamefont {Grazzini}, \citenamefont {Kallweit},\ and\ \citenamefont
  {Mazzitelli}}]{Catani:2019hip}%
  \BibitemOpen
  \bibfield  {author} {\bibinfo {author} {\bibfnamefont {S.}~\bibnamefont
  {Catani}}, \bibinfo {author} {\bibfnamefont {S.}~\bibnamefont {Devoto}},
  \bibinfo {author} {\bibfnamefont {M.}~\bibnamefont {Grazzini}}, \bibinfo
  {author} {\bibfnamefont {S.}~\bibnamefont {Kallweit}}, \ and\ \bibinfo
  {author} {\bibfnamefont {J.}~\bibnamefont {Mazzitelli}},\ }\href {\doibase
  10.1007/JHEP07(2019)100} {\bibfield  {journal} {\bibinfo  {journal} {JHEP}\
  }\textbf {\bibinfo {volume} {07}},\ \bibinfo {pages} {100} (\bibinfo {year}
  {2019}{\natexlab{b}})},\ \Eprint {http://arxiv.org/abs/1906.06535}
  {arXiv:1906.06535 [hep-ph]} \BibitemShut {NoStop}%
\bibitem [{\citenamefont {Beneke}\ \emph {et~al.}(2010)\citenamefont {Beneke},
  \citenamefont {Czakon}, \citenamefont {Falgari}, \citenamefont {Mitov},\ and\
  \citenamefont {Schwinn}}]{Beneke:2009ye}%
  \BibitemOpen
  \bibfield  {author} {\bibinfo {author} {\bibfnamefont {M.}~\bibnamefont
  {Beneke}}, \bibinfo {author} {\bibfnamefont {M.}~\bibnamefont {Czakon}},
  \bibinfo {author} {\bibfnamefont {P.}~\bibnamefont {Falgari}}, \bibinfo
  {author} {\bibfnamefont {A.}~\bibnamefont {Mitov}}, \ and\ \bibinfo {author}
  {\bibfnamefont {C.}~\bibnamefont {Schwinn}},\ }\href {\doibase
  10.1016/j.physletb.2010.05.038} {\bibfield  {journal} {\bibinfo  {journal}
  {Phys. Lett.}\ }\textbf {\bibinfo {volume} {B690}},\ \bibinfo {pages} {483}
  (\bibinfo {year} {2010})},\ \Eprint {http://arxiv.org/abs/0911.5166}
  {arXiv:0911.5166 [hep-ph]} \BibitemShut {NoStop}%
\bibitem [{\citenamefont {Czakon}\ \emph {et~al.}(2009)\citenamefont {Czakon},
  \citenamefont {Mitov},\ and\ \citenamefont {Sterman}}]{Czakon:2009zw}%
  \BibitemOpen
  \bibfield  {author} {\bibinfo {author} {\bibfnamefont {M.}~\bibnamefont
  {Czakon}}, \bibinfo {author} {\bibfnamefont {A.}~\bibnamefont {Mitov}}, \
  and\ \bibinfo {author} {\bibfnamefont {G.~F.}\ \bibnamefont {Sterman}},\
  }\href {\doibase 10.1103/PhysRevD.80.074017} {\bibfield  {journal} {\bibinfo
  {journal} {Phys. Rev.}\ }\textbf {\bibinfo {volume} {D80}},\ \bibinfo {pages}
  {074017} (\bibinfo {year} {2009})},\ \Eprint {http://arxiv.org/abs/0907.1790}
  {arXiv:0907.1790 [hep-ph]} \BibitemShut {NoStop}%
\bibitem [{\citenamefont {Beneke}\ \emph {et~al.}(2012)\citenamefont {Beneke},
  \citenamefont {Falgari}, \citenamefont {Klein},\ and\ \citenamefont
  {Schwinn}}]{Beneke:2011mq}%
  \BibitemOpen
  \bibfield  {author} {\bibinfo {author} {\bibfnamefont {M.}~\bibnamefont
  {Beneke}}, \bibinfo {author} {\bibfnamefont {P.}~\bibnamefont {Falgari}},
  \bibinfo {author} {\bibfnamefont {S.}~\bibnamefont {Klein}}, \ and\ \bibinfo
  {author} {\bibfnamefont {C.}~\bibnamefont {Schwinn}},\ }\href {\doibase
  10.1016/j.nuclphysb.2011.10.021} {\bibfield  {journal} {\bibinfo  {journal}
  {Nucl. Phys.}\ }\textbf {\bibinfo {volume} {B855}},\ \bibinfo {pages} {695}
  (\bibinfo {year} {2012})},\ \Eprint {http://arxiv.org/abs/1109.1536}
  {arXiv:1109.1536 [hep-ph]} \BibitemShut {NoStop}%
\bibitem [{\citenamefont {Cacciari}\ \emph {et~al.}(2012)\citenamefont
  {Cacciari}, \citenamefont {Czakon}, \citenamefont {Mangano}, \citenamefont
  {Mitov},\ and\ \citenamefont {Nason}}]{Cacciari:2011hy}%
  \BibitemOpen
  \bibfield  {author} {\bibinfo {author} {\bibfnamefont {M.}~\bibnamefont
  {Cacciari}}, \bibinfo {author} {\bibfnamefont {M.}~\bibnamefont {Czakon}},
  \bibinfo {author} {\bibfnamefont {M.}~\bibnamefont {Mangano}}, \bibinfo
  {author} {\bibfnamefont {A.}~\bibnamefont {Mitov}}, \ and\ \bibinfo {author}
  {\bibfnamefont {P.}~\bibnamefont {Nason}},\ }\href {\doibase
  10.1016/j.physletb.2012.03.013} {\bibfield  {journal} {\bibinfo  {journal}
  {Phys. Lett.}\ }\textbf {\bibinfo {volume} {B710}},\ \bibinfo {pages} {612}
  (\bibinfo {year} {2012})},\ \Eprint {http://arxiv.org/abs/1111.5869}
  {arXiv:1111.5869 [hep-ph]} \BibitemShut {NoStop}%
\bibitem [{\citenamefont {Kidonakis}(2014)}]{Kidonakis:2012rm}%
  \BibitemOpen
  \bibfield  {author} {\bibinfo {author} {\bibfnamefont {N.}~\bibnamefont
  {Kidonakis}},\ }\href {\doibase 10.1134/S1063779614040091} {\bibfield
  {journal} {\bibinfo  {journal} {Phys. Part. Nucl.}\ }\textbf {\bibinfo
  {volume} {45}},\ \bibinfo {pages} {714} (\bibinfo {year} {2014})},\ \Eprint
  {http://arxiv.org/abs/1210.7813} {arXiv:1210.7813 [hep-ph]} \BibitemShut
  {NoStop}%
\bibitem [{\citenamefont {Ferroglia}\ \emph {et~al.}(2012)\citenamefont
  {Ferroglia}, \citenamefont {Pecjak},\ and\ \citenamefont
  {Yang}}]{Ferroglia:2012ku}%
  \BibitemOpen
  \bibfield  {author} {\bibinfo {author} {\bibfnamefont {A.}~\bibnamefont
  {Ferroglia}}, \bibinfo {author} {\bibfnamefont {B.~D.}\ \bibnamefont
  {Pecjak}}, \ and\ \bibinfo {author} {\bibfnamefont {L.~L.}\ \bibnamefont
  {Yang}},\ }\href {\doibase 10.1103/PhysRevD.86.034010} {\bibfield  {journal}
  {\bibinfo  {journal} {Phys. Rev.}\ }\textbf {\bibinfo {volume} {D86}},\
  \bibinfo {pages} {034010} (\bibinfo {year} {2012})},\ \Eprint
  {http://arxiv.org/abs/1205.3662} {arXiv:1205.3662 [hep-ph]} \BibitemShut
  {NoStop}%
\bibitem [{\citenamefont {Ferroglia}\ \emph {et~al.}(2014)\citenamefont
  {Ferroglia}, \citenamefont {Marzani}, \citenamefont {Pecjak},\ and\
  \citenamefont {Yang}}]{Ferroglia:2013awa}%
  \BibitemOpen
  \bibfield  {author} {\bibinfo {author} {\bibfnamefont {A.}~\bibnamefont
  {Ferroglia}}, \bibinfo {author} {\bibfnamefont {S.}~\bibnamefont {Marzani}},
  \bibinfo {author} {\bibfnamefont {B.~D.}\ \bibnamefont {Pecjak}}, \ and\
  \bibinfo {author} {\bibfnamefont {L.~L.}\ \bibnamefont {Yang}},\ }\href
  {\doibase 10.1007/JHEP01(2014)028} {\bibfield  {journal} {\bibinfo  {journal}
  {JHEP}\ }\textbf {\bibinfo {volume} {01}},\ \bibinfo {pages} {028} (\bibinfo
  {year} {2014})},\ \Eprint {http://arxiv.org/abs/1310.3836} {arXiv:1310.3836
  [hep-ph]} \BibitemShut {NoStop}%
\bibitem [{\citenamefont {Czakon}\ \emph {et~al.}(2018)\citenamefont {Czakon},
  \citenamefont {Ferroglia}, \citenamefont {Heymes}, \citenamefont {Mitov},
  \citenamefont {Pecjak}, \citenamefont {Scott}, \citenamefont {Wang},\ and\
  \citenamefont {Yang}}]{Czakon:2018nun}%
  \BibitemOpen
  \bibfield  {author} {\bibinfo {author} {\bibfnamefont {M.}~\bibnamefont
  {Czakon}}, \bibinfo {author} {\bibfnamefont {A.}~\bibnamefont {Ferroglia}},
  \bibinfo {author} {\bibfnamefont {D.}~\bibnamefont {Heymes}}, \bibinfo
  {author} {\bibfnamefont {A.}~\bibnamefont {Mitov}}, \bibinfo {author}
  {\bibfnamefont {B.~D.}\ \bibnamefont {Pecjak}}, \bibinfo {author}
  {\bibfnamefont {D.~J.}\ \bibnamefont {Scott}}, \bibinfo {author}
  {\bibfnamefont {X.}~\bibnamefont {Wang}}, \ and\ \bibinfo {author}
  {\bibfnamefont {L.~L.}\ \bibnamefont {Yang}},\ }\href {\doibase
  10.1007/JHEP05(2018)149} {\bibfield  {journal} {\bibinfo  {journal} {JHEP}\
  }\textbf {\bibinfo {volume} {05}},\ \bibinfo {pages} {149} (\bibinfo {year}
  {2018})},\ \Eprint {http://arxiv.org/abs/1803.07623} {arXiv:1803.07623
  [hep-ph]} \BibitemShut {NoStop}%
\bibitem [{\citenamefont {Beenakker}\ \emph {et~al.}(1994)\citenamefont
  {Beenakker}, \citenamefont {Denner}, \citenamefont {Hollik}, \citenamefont
  {Mertig}, \citenamefont {Sack},\ and\ \citenamefont
  {Wackeroth}}]{Beenakker:1993yr}%
  \BibitemOpen
  \bibfield  {author} {\bibinfo {author} {\bibfnamefont {W.}~\bibnamefont
  {Beenakker}}, \bibinfo {author} {\bibfnamefont {A.}~\bibnamefont {Denner}},
  \bibinfo {author} {\bibfnamefont {W.}~\bibnamefont {Hollik}}, \bibinfo
  {author} {\bibfnamefont {R.}~\bibnamefont {Mertig}}, \bibinfo {author}
  {\bibfnamefont {T.}~\bibnamefont {Sack}}, \ and\ \bibinfo {author}
  {\bibfnamefont {D.}~\bibnamefont {Wackeroth}},\ }\href {\doibase
  10.1016/0550-3213(94)90454-5} {\bibfield  {journal} {\bibinfo  {journal}
  {Nucl. Phys.}\ }\textbf {\bibinfo {volume} {B411}},\ \bibinfo {pages} {343}
  (\bibinfo {year} {1994})}\BibitemShut {NoStop}%
\bibitem [{\citenamefont {Bernreuther}\ \emph {et~al.}(2006)\citenamefont
  {Bernreuther}, \citenamefont {Fuecker},\ and\ \citenamefont
  {Si}}]{Bernreuther:2006vg}%
  \BibitemOpen
  \bibfield  {author} {\bibinfo {author} {\bibfnamefont {W.}~\bibnamefont
  {Bernreuther}}, \bibinfo {author} {\bibfnamefont {M.}~\bibnamefont
  {Fuecker}}, \ and\ \bibinfo {author} {\bibfnamefont {Z.-G.}\ \bibnamefont
  {Si}},\ }\href {\doibase 10.1103/PhysRevD.74.113005} {\bibfield  {journal}
  {\bibinfo  {journal} {Phys. Rev.}\ }\textbf {\bibinfo {volume} {D74}},\
  \bibinfo {pages} {113005} (\bibinfo {year} {2006})},\ \Eprint
  {http://arxiv.org/abs/hep-ph/0610334} {arXiv:hep-ph/0610334 [hep-ph]}
  \BibitemShut {NoStop}%
\bibitem [{\citenamefont {Kühn}\ \emph {et~al.}(2007)\citenamefont {Kühn},
  \citenamefont {Scharf},\ and\ \citenamefont {Uwer}}]{Kuhn:2006vh}%
  \BibitemOpen
  \bibfield  {author} {\bibinfo {author} {\bibfnamefont {J.~H.}\ \bibnamefont
  {Kühn}}, \bibinfo {author} {\bibfnamefont {A.}~\bibnamefont {Scharf}}, \
  and\ \bibinfo {author} {\bibfnamefont {P.}~\bibnamefont {Uwer}},\ }\href
  {\doibase 10.1140/epjc/s10052-007-0275-x} {\bibfield  {journal} {\bibinfo
  {journal} {Eur. Phys. J.}\ }\textbf {\bibinfo {volume} {C51}},\ \bibinfo
  {pages} {37} (\bibinfo {year} {2007})},\ \Eprint
  {http://arxiv.org/abs/hep-ph/0610335} {arXiv:hep-ph/0610335 [hep-ph]}
  \BibitemShut {NoStop}%
\bibitem [{\citenamefont {Moretti}\ \emph {et~al.}(2006)\citenamefont
  {Moretti}, \citenamefont {Nolten},\ and\ \citenamefont
  {Ross}}]{Moretti:2006nf}%
  \BibitemOpen
  \bibfield  {author} {\bibinfo {author} {\bibfnamefont {S.}~\bibnamefont
  {Moretti}}, \bibinfo {author} {\bibfnamefont {M.~R.}\ \bibnamefont {Nolten}},
  \ and\ \bibinfo {author} {\bibfnamefont {D.~A.}\ \bibnamefont {Ross}},\
  }\href {\doibase 10.1016/j.physletb.2006.06.078,
  10.1016/j.physletb.2007.10.039} {\bibfield  {journal} {\bibinfo  {journal}
  {Phys. Lett.}\ }\textbf {\bibinfo {volume} {B639}},\ \bibinfo {pages} {513}
  (\bibinfo {year} {2006})},\ \bibinfo {note} {[Erratum: Phys.
  Lett.B660,607(2008)]},\ \Eprint {http://arxiv.org/abs/hep-ph/0603083}
  {arXiv:hep-ph/0603083 [hep-ph]} \BibitemShut {NoStop}%
\bibitem [{\citenamefont {Pagani}\ \emph {et~al.}(2016)\citenamefont {Pagani},
  \citenamefont {Tsinikos},\ and\ \citenamefont {Zaro}}]{Pagani:2016caq}%
  \BibitemOpen
  \bibfield  {author} {\bibinfo {author} {\bibfnamefont {D.}~\bibnamefont
  {Pagani}}, \bibinfo {author} {\bibfnamefont {I.}~\bibnamefont {Tsinikos}}, \
  and\ \bibinfo {author} {\bibfnamefont {M.}~\bibnamefont {Zaro}},\ }\href
  {\doibase 10.1140/epjc/s10052-016-4318-z} {\bibfield  {journal} {\bibinfo
  {journal} {Eur. Phys. J.}\ }\textbf {\bibinfo {volume} {C76}},\ \bibinfo
  {pages} {479} (\bibinfo {year} {2016})},\ \Eprint
  {http://arxiv.org/abs/1606.01915} {arXiv:1606.01915 [hep-ph]} \BibitemShut
  {NoStop}%
\bibitem [{\citenamefont {Martini}\ and\ \citenamefont
  {Uwer}(2015)}]{Martini:2015fsa}%
  \BibitemOpen
  \bibfield  {author} {\bibinfo {author} {\bibfnamefont {T.}~\bibnamefont
  {Martini}}\ and\ \bibinfo {author} {\bibfnamefont {P.}~\bibnamefont {Uwer}},\
  }\href {\doibase 10.1007/JHEP09(2015)083} {\bibfield  {journal} {\bibinfo
  {journal} {JHEP}\ }\textbf {\bibinfo {volume} {09}},\ \bibinfo {pages} {083}
  (\bibinfo {year} {2015})},\ \Eprint {http://arxiv.org/abs/1506.08798}
  {arXiv:1506.08798 [hep-ph]} \BibitemShut {NoStop}%
\bibitem [{\citenamefont {Martini}\ and\ \citenamefont
  {Uwer}(2018)}]{Martini:2017ydu}%
  \BibitemOpen
  \bibfield  {author} {\bibinfo {author} {\bibfnamefont {T.}~\bibnamefont
  {Martini}}\ and\ \bibinfo {author} {\bibfnamefont {P.}~\bibnamefont {Uwer}},\
  }\href {\doibase 10.1007/JHEP05(2018)141} {\bibfield  {journal} {\bibinfo
  {journal} {JHEP}\ }\textbf {\bibinfo {volume} {05}},\ \bibinfo {pages} {141}
  (\bibinfo {year} {2018})},\ \Eprint {http://arxiv.org/abs/1712.04527}
  {arXiv:1712.04527 [hep-ph]} \BibitemShut {NoStop}%
\bibitem [{\citenamefont {Kraus}\ \emph
  {et~al.}(2019{\natexlab{a}})\citenamefont {Kraus}, \citenamefont {Martini},\
  and\ \citenamefont {Uwer}}]{Kraus:2019qoq}%
  \BibitemOpen
  \bibfield  {author} {\bibinfo {author} {\bibfnamefont {M.}~\bibnamefont
  {Kraus}}, \bibinfo {author} {\bibfnamefont {T.}~\bibnamefont {Martini}}, \
  and\ \bibinfo {author} {\bibfnamefont {P.}~\bibnamefont {Uwer}},\ }\href
  {\doibase 10.1103/PhysRevD.100.076010} {\bibfield  {journal} {\bibinfo
  {journal} {Phys. Rev. D}\ }\textbf {\bibinfo {volume} {100}},\ \bibinfo
  {pages} {076010} (\bibinfo {year} {2019}{\natexlab{a}})},\ \Eprint
  {http://arxiv.org/abs/1901.08008} {arXiv:1901.08008 [hep-ph]} \BibitemShut
  {NoStop}%
\bibitem [{\citenamefont {Kraus}\ \emph
  {et~al.}(2019{\natexlab{b}})\citenamefont {Kraus}, \citenamefont {Martini},
  \citenamefont {Peitzsch},\ and\ \citenamefont {Uwer}}]{Kraus:2019myc}%
  \BibitemOpen
  \bibfield  {author} {\bibinfo {author} {\bibfnamefont {M.}~\bibnamefont
  {Kraus}}, \bibinfo {author} {\bibfnamefont {T.}~\bibnamefont {Martini}},
  \bibinfo {author} {\bibfnamefont {S.}~\bibnamefont {Peitzsch}}, \ and\
  \bibinfo {author} {\bibfnamefont {P.}~\bibnamefont {Uwer}},\ }\href@noop {}
  {\  (\bibinfo {year} {2019}{\natexlab{b}})},\ \Eprint
  {http://arxiv.org/abs/1908.09100} {arXiv:1908.09100 [hep-ph]} \BibitemShut
  {NoStop}%
\bibitem [{\citenamefont {Nason}(2004)}]{Nason:2004rx}%
  \BibitemOpen
  \bibfield  {author} {\bibinfo {author} {\bibfnamefont {P.}~\bibnamefont
  {Nason}},\ }\href {\doibase 10.1088/1126-6708/2004/11/040} {\bibfield
  {journal} {\bibinfo  {journal} {JHEP}\ }\textbf {\bibinfo {volume} {11}},\
  \bibinfo {pages} {040} (\bibinfo {year} {2004})},\ \Eprint
  {http://arxiv.org/abs/hep-ph/0409146} {arXiv:hep-ph/0409146} \BibitemShut
  {NoStop}%
\bibitem [{\citenamefont {Sjostrand}\ \emph {et~al.}(2006)\citenamefont
  {Sjostrand}, \citenamefont {Mrenna},\ and\ \citenamefont
  {Skands}}]{Sjostrand:2006za}%
  \BibitemOpen
  \bibfield  {author} {\bibinfo {author} {\bibfnamefont {T.}~\bibnamefont
  {Sjostrand}}, \bibinfo {author} {\bibfnamefont {S.}~\bibnamefont {Mrenna}}, \
  and\ \bibinfo {author} {\bibfnamefont {P.~Z.}\ \bibnamefont {Skands}},\
  }\href {\doibase 10.1088/1126-6708/2006/05/026} {\bibfield  {journal}
  {\bibinfo  {journal} {JHEP}\ }\textbf {\bibinfo {volume} {05}},\ \bibinfo
  {pages} {026} (\bibinfo {year} {2006})},\ \Eprint
  {http://arxiv.org/abs/hep-ph/0603175} {arXiv:hep-ph/0603175} \BibitemShut
  {NoStop}%
\bibitem [{\citenamefont {Frixione}\ \emph
  {et~al.}(2007{\natexlab{a}})\citenamefont {Frixione}, \citenamefont {Nason},\
  and\ \citenamefont {Oleari}}]{Frixione:2007vw}%
  \BibitemOpen
  \bibfield  {author} {\bibinfo {author} {\bibfnamefont {S.}~\bibnamefont
  {Frixione}}, \bibinfo {author} {\bibfnamefont {P.}~\bibnamefont {Nason}}, \
  and\ \bibinfo {author} {\bibfnamefont {C.}~\bibnamefont {Oleari}},\ }\href
  {\doibase 10.1088/1126-6708/2007/11/070} {\bibfield  {journal} {\bibinfo
  {journal} {JHEP}\ }\textbf {\bibinfo {volume} {11}},\ \bibinfo {pages} {070}
  (\bibinfo {year} {2007}{\natexlab{a}})},\ \Eprint
  {http://arxiv.org/abs/0709.2092} {arXiv:0709.2092 [hep-ph]} \BibitemShut
  {NoStop}%
\bibitem [{\citenamefont {Frixione}\ \emph
  {et~al.}(2007{\natexlab{b}})\citenamefont {Frixione}, \citenamefont {Nason},\
  and\ \citenamefont {Ridolfi}}]{Frixione:2007nw}%
  \BibitemOpen
  \bibfield  {author} {\bibinfo {author} {\bibfnamefont {S.}~\bibnamefont
  {Frixione}}, \bibinfo {author} {\bibfnamefont {P.}~\bibnamefont {Nason}}, \
  and\ \bibinfo {author} {\bibfnamefont {G.}~\bibnamefont {Ridolfi}},\ }\href
  {\doibase 10.1088/1126-6708/2007/09/126} {\bibfield  {journal} {\bibinfo
  {journal} {JHEP}\ }\textbf {\bibinfo {volume} {09}},\ \bibinfo {pages} {126}
  (\bibinfo {year} {2007}{\natexlab{b}})},\ \Eprint
  {http://arxiv.org/abs/0707.3088} {arXiv:0707.3088 [hep-ph]} \BibitemShut
  {NoStop}%
\bibitem [{\citenamefont {Alioli}\ \emph {et~al.}(2010)\citenamefont {Alioli},
  \citenamefont {Nason}, \citenamefont {Oleari},\ and\ \citenamefont
  {Re}}]{Alioli:2010xd}%
  \BibitemOpen
  \bibfield  {author} {\bibinfo {author} {\bibfnamefont {S.}~\bibnamefont
  {Alioli}}, \bibinfo {author} {\bibfnamefont {P.}~\bibnamefont {Nason}},
  \bibinfo {author} {\bibfnamefont {C.}~\bibnamefont {Oleari}}, \ and\ \bibinfo
  {author} {\bibfnamefont {E.}~\bibnamefont {Re}},\ }\href {\doibase
  10.1007/JHEP06(2010)043} {\bibfield  {journal} {\bibinfo  {journal} {JHEP}\
  }\textbf {\bibinfo {volume} {06}},\ \bibinfo {pages} {043} (\bibinfo {year}
  {2010})},\ \Eprint {http://arxiv.org/abs/1002.2581} {arXiv:1002.2581
  [hep-ph]} \BibitemShut {NoStop}%
\bibitem [{\citenamefont {Giele}\ \emph {et~al.}(1993)\citenamefont {Giele},
  \citenamefont {Glover},\ and\ \citenamefont {Kosower}}]{Giele:1993dj}%
  \BibitemOpen
  \bibfield  {author} {\bibinfo {author} {\bibfnamefont {W.~T.}\ \bibnamefont
  {Giele}}, \bibinfo {author} {\bibfnamefont {E.~W.~N.}\ \bibnamefont
  {Glover}}, \ and\ \bibinfo {author} {\bibfnamefont {D.~A.}\ \bibnamefont
  {Kosower}},\ }\href {\doibase 10.1016/0550-3213(93)90365-V} {\bibfield
  {journal} {\bibinfo  {journal} {Nucl. Phys.}\ }\textbf {\bibinfo {volume}
  {B403}},\ \bibinfo {pages} {633} (\bibinfo {year} {1993})},\ \Eprint
  {http://arxiv.org/abs/hep-ph/9302225} {arXiv:hep-ph/9302225 [hep-ph]}
  \BibitemShut {NoStop}%
\bibitem [{\citenamefont {Badger}\ \emph {et~al.}(2011)\citenamefont {Badger},
  \citenamefont {Sattler},\ and\ \citenamefont {Yundin}}]{Badger:2011yu}%
  \BibitemOpen
  \bibfield  {author} {\bibinfo {author} {\bibfnamefont {S.}~\bibnamefont
  {Badger}}, \bibinfo {author} {\bibfnamefont {R.}~\bibnamefont {Sattler}}, \
  and\ \bibinfo {author} {\bibfnamefont {V.}~\bibnamefont {Yundin}},\ }\href
  {\doibase 10.1103/PhysRevD.83.074020} {\bibfield  {journal} {\bibinfo
  {journal} {Phys. Rev. D}\ }\textbf {\bibinfo {volume} {83}},\ \bibinfo
  {pages} {074020} (\bibinfo {year} {2011})},\ \Eprint
  {http://arxiv.org/abs/1101.5947} {arXiv:1101.5947 [hep-ph]} \BibitemShut
  {NoStop}%
\bibitem [{\citenamefont {Kinoshita}(1962)}]{Kinoshita:1962ur}%
  \BibitemOpen
  \bibfield  {author} {\bibinfo {author} {\bibfnamefont {T.}~\bibnamefont
  {Kinoshita}},\ }\href {\doibase 10.1063/1.1724268} {\bibfield  {journal}
  {\bibinfo  {journal} {J. Math. Phys.}\ }\textbf {\bibinfo {volume} {3}},\
  \bibinfo {pages} {650} (\bibinfo {year} {1962})}\BibitemShut {NoStop}%
\bibitem [{\citenamefont {Lee}\ and\ \citenamefont
  {Nauenberg}(1964)}]{Lee:1964is}%
  \BibitemOpen
  \bibfield  {author} {\bibinfo {author} {\bibfnamefont {T.~D.}\ \bibnamefont
  {Lee}}\ and\ \bibinfo {author} {\bibfnamefont {M.}~\bibnamefont
  {Nauenberg}},\ }\href {\doibase 10.1103/PhysRev.133.B1549} {\bibfield
  {journal} {\bibinfo  {journal} {Phys. Rev.}\ }\textbf {\bibinfo {volume}
  {133}},\ \bibinfo {pages} {B1549} (\bibinfo {year} {1964})}\BibitemShut
  {NoStop}%
\bibitem [{\citenamefont {Aliev}\ \emph {et~al.}(2011)\citenamefont {Aliev},
  \citenamefont {Lacker}, \citenamefont {Langenfeld}, \citenamefont {Moch},
  \citenamefont {Uwer} \emph {et~al.}}]{Aliev:2010zk}%
  \BibitemOpen
  \bibfield  {author} {\bibinfo {author} {\bibfnamefont {M.}~\bibnamefont
  {Aliev}}, \bibinfo {author} {\bibfnamefont {H.}~\bibnamefont {Lacker}},
  \bibinfo {author} {\bibfnamefont {U.}~\bibnamefont {Langenfeld}}, \bibinfo
  {author} {\bibfnamefont {S.}~\bibnamefont {Moch}}, \bibinfo {author}
  {\bibfnamefont {P.}~\bibnamefont {Uwer}},  \emph {et~al.},\ }\href {\doibase
  10.1016/j.cpc.2010.12.040} {\bibfield  {journal} {\bibinfo  {journal}
  {Comput.Phys.Commun.}\ }\textbf {\bibinfo {volume} {182}},\ \bibinfo {pages}
  {1034} (\bibinfo {year} {2011})},\ \Eprint {http://arxiv.org/abs/1007.1327}
  {arXiv:1007.1327 [hep-ph]} \BibitemShut {NoStop}%
\bibitem [{\citenamefont {Alwall}\ \emph {et~al.}(2014)\citenamefont {Alwall},
  \citenamefont {Frederix}, \citenamefont {Frixione}, \citenamefont {Hirschi},
  \citenamefont {Maltoni}, \citenamefont {Mattelaer}, \citenamefont {Shao},
  \citenamefont {Stelzer}, \citenamefont {Torrielli},\ and\ \citenamefont
  {Zaro}}]{Alwall:2014hca}%
  \BibitemOpen
  \bibfield  {author} {\bibinfo {author} {\bibfnamefont {J.}~\bibnamefont
  {Alwall}}, \bibinfo {author} {\bibfnamefont {R.}~\bibnamefont {Frederix}},
  \bibinfo {author} {\bibfnamefont {S.}~\bibnamefont {Frixione}}, \bibinfo
  {author} {\bibfnamefont {V.}~\bibnamefont {Hirschi}}, \bibinfo {author}
  {\bibfnamefont {F.}~\bibnamefont {Maltoni}}, \bibinfo {author} {\bibfnamefont
  {O.}~\bibnamefont {Mattelaer}}, \bibinfo {author} {\bibfnamefont {H.~S.}\
  \bibnamefont {Shao}}, \bibinfo {author} {\bibfnamefont {T.}~\bibnamefont
  {Stelzer}}, \bibinfo {author} {\bibfnamefont {P.}~\bibnamefont {Torrielli}},
  \ and\ \bibinfo {author} {\bibfnamefont {M.}~\bibnamefont {Zaro}},\ }\href
  {\doibase 10.1007/JHEP07(2014)079} {\bibfield  {journal} {\bibinfo  {journal}
  {JHEP}\ }\textbf {\bibinfo {volume} {07}},\ \bibinfo {pages} {079} (\bibinfo
  {year} {2014})},\ \Eprint {http://arxiv.org/abs/1405.0301} {arXiv:1405.0301
  [hep-ph]} \BibitemShut {NoStop}%
\bibitem [{\citenamefont {von Neumann}(1951)}]{vonNeumann1951}%
  \BibitemOpen
  \bibfield  {author} {\bibinfo {author} {\bibfnamefont {J.}~\bibnamefont {von
  Neumann}},\ }in\ \href@noop {} {\emph {\bibinfo {booktitle} {Monte Carlo
  Method}}},\ \bibinfo {series} {National Bureau of Standards Applied
  Mathematics Series}, Vol.~\bibinfo {volume} {12},\ \bibinfo {editor} {edited
  by\ \bibinfo {editor} {\bibfnamefont {A.~S.}\ \bibnamefont {Householder}},
  \bibinfo {editor} {\bibfnamefont {G.~E.}\ \bibnamefont {Forsythe}}, \ and\
  \bibinfo {editor} {\bibfnamefont {H.~H.}\ \bibnamefont {Germond}}}\ (\bibinfo
   {publisher} {US Government Printing Office},\ \bibinfo {address}
  {Washington, DC},\ \bibinfo {year} {1951})\ Chap.~\bibinfo {chapter} {13},
  pp.\ \bibinfo {pages} {36--38}\BibitemShut {NoStop}%
\bibitem [{\citenamefont {Butter}\ \emph {et~al.}(2022)\citenamefont {Butter},
  \citenamefont {Heimel}, \citenamefont {Martini}, \citenamefont {Peitzsch},\
  and\ \citenamefont {Plehn}}]{Butter:2022vkj}%
  \BibitemOpen
  \bibfield  {author} {\bibinfo {author} {\bibfnamefont {A.}~\bibnamefont
  {Butter}}, \bibinfo {author} {\bibfnamefont {T.}~\bibnamefont {Heimel}},
  \bibinfo {author} {\bibfnamefont {T.}~\bibnamefont {Martini}}, \bibinfo
  {author} {\bibfnamefont {S.}~\bibnamefont {Peitzsch}}, \ and\ \bibinfo
  {author} {\bibfnamefont {T.}~\bibnamefont {Plehn}},\ }\href@noop {} {\
  (\bibinfo {year} {2022})},\ \Eprint {http://arxiv.org/abs/2210.00019}
  {arXiv:2210.00019 [hep-ph]} \BibitemShut {NoStop}%
\bibitem [{\citenamefont {Catani}\ and\ \citenamefont
  {Seymour}(1997)}]{Catani:1996vz}%
  \BibitemOpen
  \bibfield  {author} {\bibinfo {author} {\bibfnamefont {S.}~\bibnamefont
  {Catani}}\ and\ \bibinfo {author} {\bibfnamefont {M.}~\bibnamefont
  {Seymour}},\ }\href {\doibase 10.1016/S0550-3213(96)00589-5} {\bibfield
  {journal} {\bibinfo  {journal} {Nucl.Phys.}\ }\textbf {\bibinfo {volume}
  {B485}},\ \bibinfo {pages} {291} (\bibinfo {year} {1997})},\ \Eprint
  {http://arxiv.org/abs/hep-ph/9605323} {arXiv:hep-ph/9605323 [hep-ph]}
  \BibitemShut {NoStop}%
\bibitem [{\citenamefont {Catani}\ \emph {et~al.}(2002)\citenamefont {Catani},
  \citenamefont {Dittmaier}, \citenamefont {Seymour},\ and\ \citenamefont
  {Trocsanyi}}]{Catani:2002hc}%
  \BibitemOpen
  \bibfield  {author} {\bibinfo {author} {\bibfnamefont {S.}~\bibnamefont
  {Catani}}, \bibinfo {author} {\bibfnamefont {S.}~\bibnamefont {Dittmaier}},
  \bibinfo {author} {\bibfnamefont {M.~H.}\ \bibnamefont {Seymour}}, \ and\
  \bibinfo {author} {\bibfnamefont {Z.}~\bibnamefont {Trocsanyi}},\ }\href
  {\doibase 10.1016/S0550-3213(02)00098-6} {\bibfield  {journal} {\bibinfo
  {journal} {Nucl.Phys.}\ }\textbf {\bibinfo {volume} {B627}},\ \bibinfo
  {pages} {189} (\bibinfo {year} {2002})},\ \Eprint
  {http://arxiv.org/abs/hep-ph/0201036} {arXiv:hep-ph/0201036 [hep-ph]}
  \BibitemShut {NoStop}%
\end{thebibliography}%
\end{document}